\documentclass[aps,pre,twocolumn,groupedaddress,superscriptaddress,showpacs]{revtex4-1}

\usepackage{amsmath}
\usepackage{amssymb}
\usepackage{graphicx}
\usepackage{float}
\usepackage{color}

\begin{document}

  \title{Adsorbed films of three-patch colloids: Continuous and discontinuous transitions between thick and
thin films}

  \author{C. S. Dias}
   \email{csdias@fc.ul.pt}
    \affiliation{Departamento de F\'{\i}sica, Faculdade de Ci\^{e}ncias, Universidade de Lisboa, P-1749-016 Lisboa, Portugal, and Centro de F\'isica Te\'orica e Computacional, Universidade de Lisboa, Avenida Professor Gama Pinto 2, P-1649-003 Lisboa, Portugal}

  \author{N. A. M. Ara\'ujo}
   \email{nmaraujo@fc.ul.pt}
   \affiliation{Departamento de F\'{\i}sica, Faculdade de Ci\^{e}ncias, Universidade de Lisboa, P-1749-016 Lisboa, Portugal, and Centro de F\'isica Te\'orica e Computacional, Universidade de Lisboa, Avenida Professor Gama Pinto 2, P-1649-003 Lisboa, Portugal}

  \author{M. M. Telo da Gama}
   \email{margarid@cii.fc.ul.pt}
    \affiliation{Departamento de F\'{\i}sica, Faculdade de Ci\^{e}ncias, Universidade de Lisboa, P-1749-016 Lisboa, Portugal, and Centro de F\'isica Te\'orica e Computacional, Universidade de Lisboa, Avenida Professor Gama Pinto 2, P-1649-003 Lisboa, Portugal}


\begin{abstract}
 We investigate numerically the role of spatial arrangement of the patches on
the irreversible adsorption of patchy colloids on a substrate. We
consider spherical three-patch colloids and study the dependence of the
kinetics on the opening angle between patches. We show that growth is
suppressed below and above minimum and maximum opening angles,
revealing two absorbing phase transitions between thick and thin film regimes. While the transition at the
minimum angle is continuous, in the Directed Percolation
class, that at the maximum angle is clearly
discontinuous. For intermediate values of the opening angle, a rough colloidal
network in the Kardar-Parisi-Zhang
universality class grows indefinitely. The nature of the transitions was analyzed 
in detail by considering bond flexibility, defined as the dispersion
of the angle between the bond and the center of the patch. For the 
range of flexibilities considered we always observe two phase
transitions. However, the range of opening angles where growth is
sustained increases with flexibility. At a tricritical flexibility, the
discontinuous transition becomes continuous. The practical implications
of our findings and the relation to other nonequilibrium transitions are
discussed.
\end{abstract}

  \maketitle

\section{Introduction}
The promise of control, through the colloidal valence, the local arrangements
of colloidal networks has posed patchy colloids under the
spotlight~\cite{Blaaderen2006,Bianchi2011,Pawar2010,Kretzschmar2011,Grzelczak2010,Wilner2012,Yi2013}.
Due to the highly directional colloid-colloid
interaction~\cite{Wang2012,Bianchi2006,Russo2009} and the possibility of
combining different patch
types~\cite{Tavares2009,Tavares2009a,DelasHeras2011,Hu2012,Angioletti-Uberti2012,Kumar2013},
the equilibrium phase diagrams are
colorful~\cite{Russo2011a,DelasHeras2012,Reinhardt2013,Smallenburg2013},
yielding a seemingly endless list of new features of practical
interest~\cite{Matthews2012,Smallenburg2013a,Coluzza2013}. 

The quest for the feasibility of the equilibrium structures has shifted
the emphasis to the
kinetics~\cite{Sciortino2009,Corezzi2009,Corezzi2012,Vasilyev2013} 
in particular to the adsorption on substrates \cite{Gnan2012,Bernardino2012,Dias2013,Dias2013a,Dias2013b}.
Substrates simultaneously improve the control over
assembly~\cite{Shyr2008,Pawar2008,Tian2010,Cadilhe2007,Araujo2008} and
provide an identifiable growth direction, which helps to characterize the
time evolution of growth and to develop strategies to obtain
heterogeneous materials~\cite{Einstein2010}. The ultimate goal 
is to combine flat or templated substrates and tunable
patchy colloids to fashion a new family of metamaterials.

\begin{figure}[t]
  \begin{center}
   \includegraphics[width=0.5\columnwidth]{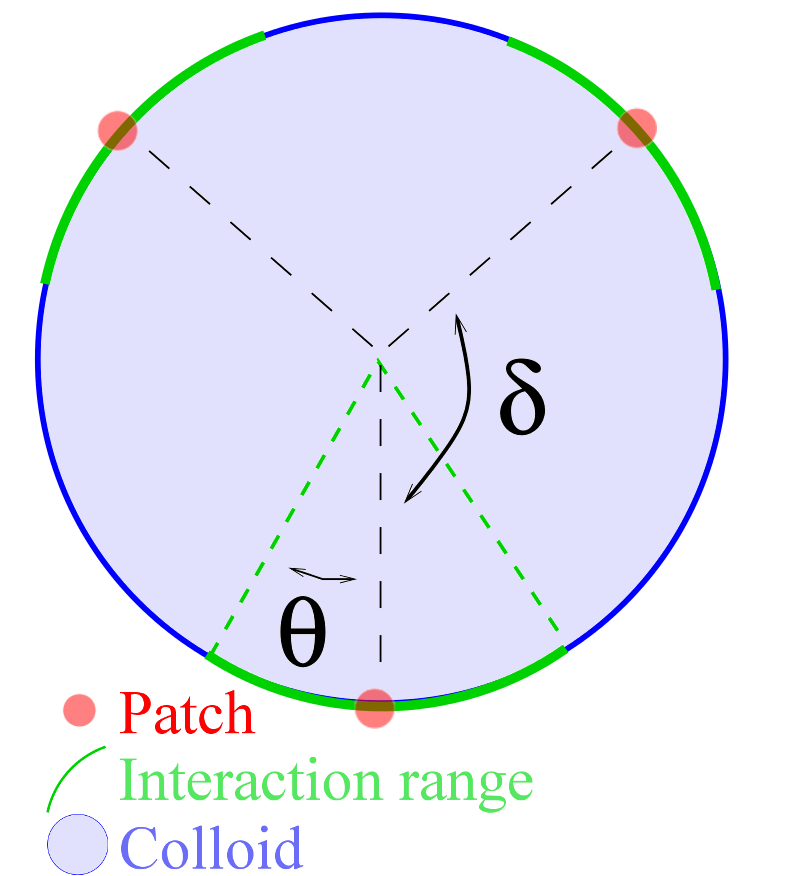} \\
   \end{center}
   \caption{(color online) Schematic representation of patches (red) on the surface
of a three-patch colloid (blue) and their interaction range $\theta$
(green). The distribution of patches is described by an opening angle
$\delta$, in units of $\pi\text{ rad}$, from the center of the
two patches and the center of the reference one. The (red) patch is the 
bonding site and its interaction range (green) represents the extent of the 
attractive interaction between patches. \label{fig.model}}
 \end{figure}

The focus of the theoretical and experimental work has been on the directionality of the interactions with the role of the patch spatial 
arrangement largely overlooked. However, recent theoretical \cite{Doppelbauer2010,Marshall2013,Marshall2013a,Tavares2014}
and experimental \cite{Iwashita2013,Iwashita2014} studies have revealed 
a strong dependence of the equilibrium structures of patchy colloids on the valence and strength of the interactions. 
Here, as a first step to understand the role of patch-patch correlations on the kinetics of aggregation, we consider 
the limit of irreversible adsorption  with  advective mass transport towards the substrate. To access large-length and 
long-time scales, we choose not to perform detailed molecular dynamics simulations and use, instead, a stochastic 
model previously proposed in Ref.~\cite{Dias2013}. As schematically represented in Fig.~\ref{fig.model} we consider three-patch spherical 
colloids and characterize the patch arrangement by the opening angle $\delta$ between a reference patch and 
the other two (adjustable patches). We found a strong dependence of the kinetics on $\delta$. In particular, 
sustained growth of a colloidal network is only possible for a finite range of opening angles $\delta$, 
above $\delta_{min}$ and below  $\delta_\mathrm{max}$. We show that the approach to these thresholds can be 
described as transitions to absorbing states, driven by different mechanisms and of different nature. 
While the transition at $\delta_\mathrm{min}$ is continuous that at $\delta_\mathrm{max}$ may be discontinuous.

In the following section we give a description of the model. In Sec.~\ref{sec.results}, we report the results in three subsections: 
A the transition at the minimum opening angle; B the transition at the maximum opening angle; 
and C the effect of bond flexibility. Finally, in Sec.~\ref{sec.conc}, we draw some conclusions.

\section{Model}
We consider spherical three-patch colloids of
unit diameter $\sigma$ and a two-dimensional system with a flat
substrate at height $h=0$. We also define $h_\mathrm{max}$ as the
maximum height of a colloid in the network and assume an initially empty
substrate, such that $h_\mathrm{max}=0$. To describe the advective
transport, we iteratively generate a horizontal position uniformly at
random at a height $h_\mathrm{dep}=h_\mathrm{max}+\sigma$ and simulate
the ballistic downward movement until the colloid
either hits the substrate or another colloid. The colloid-substrate
collision always results on adsorption of the colloid with a random
orientation.

The patch-patch short-range interaction is described in a stochastic way as first proposed 
in Ref.~\cite{Dias2013}. We focus on chemical or DNA mediated bonds \cite{Geerts2010,Wang2012}, which 
are highly directional and very strong, and may be considered irreversible within the timescale of interest \cite{Leunissen2011}.
Thus, we assume that two patches bond in an irreversible way, a process we name binding, and that bonds are optimal such 
that the center of two bonded colloids is always aligned with their bonding patches.
We define for each patch an interaction range around the patch, represented by the thick (green)
line in Fig.~\ref{fig.model}, which accounts for both the extension
of the patch and the range of the patch-patch interaction. The interaction
range is characterized by a single parameter $\theta=\pi/6$,
representing the maximum angle with the center of the patch (see
Fig.~\ref{fig.model}). Two patches may bind if their interaction ranges
partially overlap in the event of a collision. Thus, stochastically,
when the incoming colloid hits the interaction range of a colloid in
the network, it binds irreversibly to it with a probability $p=A_\mathrm{ir}/A$, where $A=\pi\sigma$ is the surface 
of the colloid and $A_\mathrm{ir}$ is the extension of the surface covered by the interaction range of all patches. 
In the case of successful binding, the binding patch of the incoming colloid is chosen uniformly at 
random among its three patches and its position and orientation is adjusted accordingly. 
Since the network colloid position and orientation are assumed irreversibly fixed, 
the alignment of the new binding patches results solely from the rotation and translation of the incoming colloid.

\section{Results}\label{sec.results}
We performed simulations for different opening angles
$\delta$ (in units of $\pi\text{ rad}$) 
and lengths of the substrate $L$
(in units of the colloid diameter). For $\delta<\delta_\mathrm{min}$, the
angle between patches is such that all patches are in the same
hemisphere. Colloids in the network will most likely have all
patches towards the substrate and incoming colloids will fail to bind. 
Thus, when no more colloids can adsorb on the substrate and after
a handful of patch-patch bindings, the growth is suppressed. For
$\delta>\delta_\mathrm{max}$, the two adjustable patches are so close
that only one can effectively bind due to the excluded volume
interaction, i.\ e., when one colloid binds to one of the adjustable
patches it inevitably shields the access of a new colloid to the second
adjustable patch.  This also hinders growth due to a more subtle
mechanism. Since only one of the adjustable patches can bind, branching
is suppressed and only linear colloidal chains grow out of the
substrate. For $\delta_\mathrm{max}<\delta<1$ these chains are locally
tilted and the growth direction fluctuates around the vertical
direction. As it fluctuates, the orientation of the patches at the tip
will eventually point down and the growth of the chain will be
suppressed. Since binding is irreversible and occurs only when an incoming colloid joins the 
network, the total number of bonds is equal to the number of colloids in the network. 
The absorbing state occurs when no more patches are available to bind incoming colloids.

\begin{figure}[t]
 \begin{center}
 \includegraphics[width=\columnwidth]{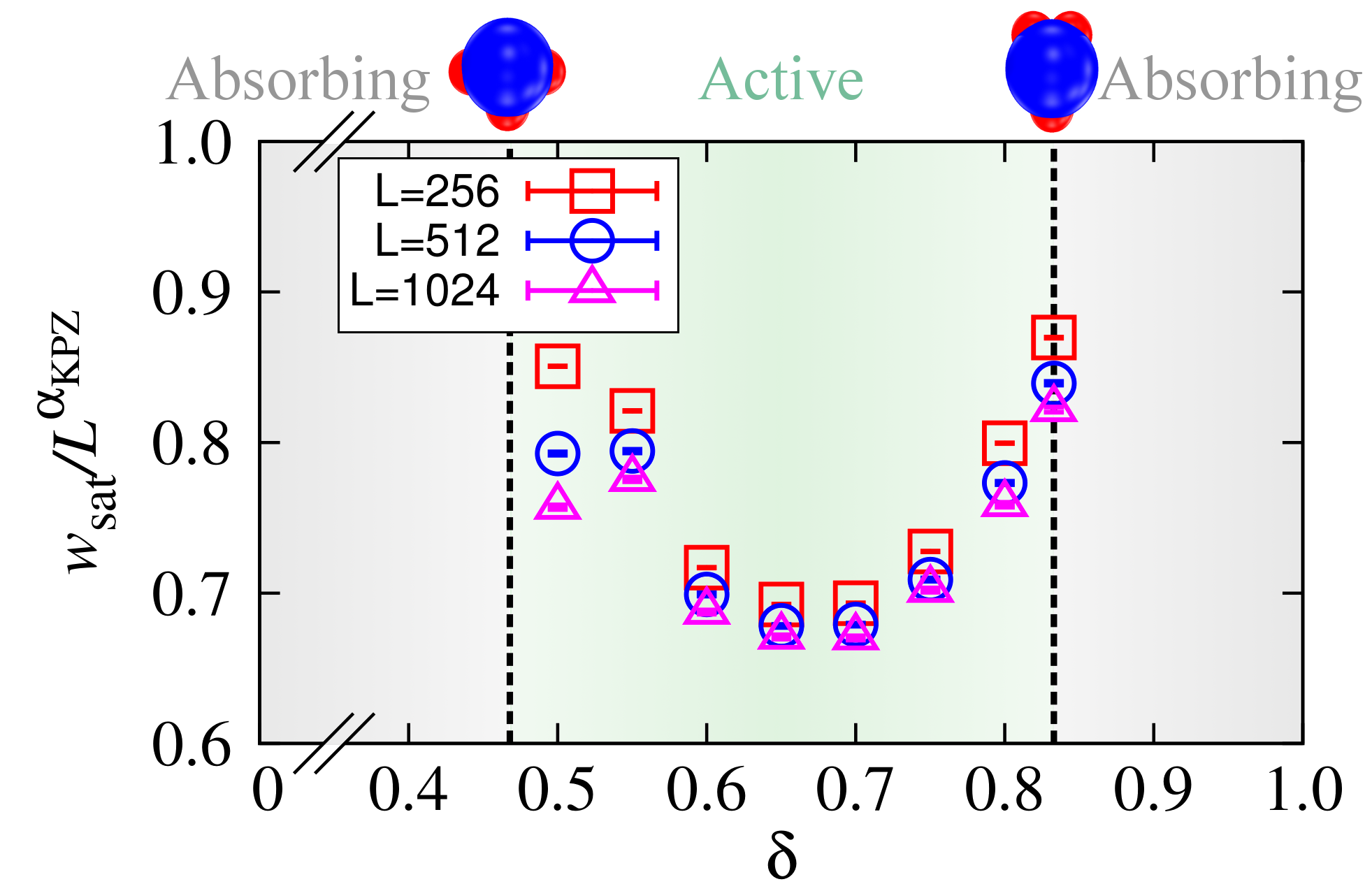} \\
 \end{center}
\caption{(color online) Adsorption of three-patch colloids for different values of the
opening angle ($\delta$). Two absorbing phases are found for
$\delta<\delta_\mathrm{min}$ and $\delta>\delta_\mathrm{max}$, where
growth is suppressed at a finite thickness. For
$\delta_\mathrm{min}<\delta<\delta_\mathrm{max}$, a sustained growth is
observed (active phase). The data points are for the data collapse of
the roughness in the active phase using
$w_\mathrm{sat}=L^{\alpha_\mathrm{KPZ}}\mathcal{F}\left[\delta\right]$,
where $\mathcal{F}$ is a scaling function and $\alpha_\mathrm{KPZ}$ is
the roughness exponent for the Kardar-Parisi-Zhang universality class.
We considered three different lengths of the substrate
$L=\{256,512,1024\}$ and results are averages over $\{320000,80000,40000\}$
samples.  \label{fig.roughness_angle} }
\end{figure}
\begin{figure*}[t]
 \begin{center}
 \includegraphics[width=2\columnwidth]{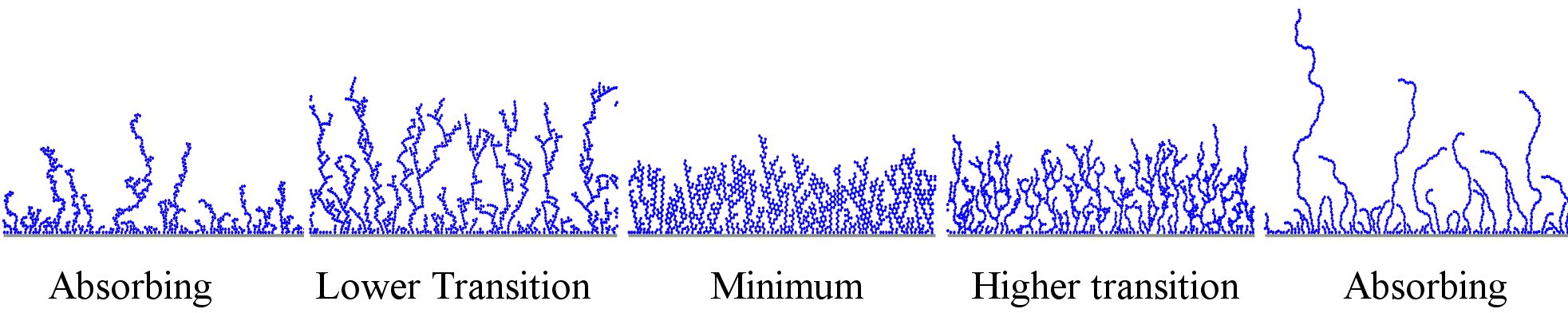} \\
 \end{center}
\caption{(color online) Snapshots for different regimes. 
From left to right $\delta=\{0.4,0.468,0.666,0.833,0.85\}$.
For a system size $L=128$ and $10L$ deposited colloids.\label{fig.snaps} }
\end{figure*}
For $\delta_\mathrm{min}<\delta<\delta_\mathrm{max}$, a ramified network
of patchy colloids grows from the substrate in a sustained fashion. To
characterize this growth, we calculate the roughness $w$ of the
interface in the following way. We divide the system in $N$ vertical
slices of width $\sigma$ ($N=L/\sigma$). For each slice $i$ we simulate
the downward trajectory of a probe colloid released from the center of
the slice at $h_\mathrm{dep}$ and calculate the height $h_i$ at which it
first touches either one colloid in the network or the substrate. The
roughness is then defined as,
\begin{equation}\label{eq.roughness}
w=\sqrt{\frac{1}{N}\sum_i{\left[h_i-\bar{h}\right]^2}} \ \ ,
\end{equation}
where $\bar{h}=\sum_i{h_i}/N$. For all system sizes, the roughness
initially increases with the number of colloids and saturates 
at $w_\mathrm{sat}$, which depends on $\delta$ and $L$.
Figure~\ref{fig.roughness_angle} shows $w_\mathrm{sat}$ as a function of
$\delta$ for different $L$. A non-monotonic dependence on $\delta$ is
observed with a minimum at $\delta\approx2/3$. This minimum occurs when the three patches are equidistant, 
which favors branching and consequently leads to a decrease of the roughness. Snapshots of the colloidal network
for different regimes can be seen in Fig.~\ref{fig.snaps}. When $w_\mathrm{sat}$
is rescaled by $L^{\alpha_\mathrm{KPZ}}$, where
$\alpha_\mathrm{KPZ}=1/2$ is the roughness exponent for the
Kardar-Parisi-Zhang universality class~\cite{Kardar1986}, 
data collapse is observed, consistent with this universality class. This result is in contrast to previous experimental 
results for spherical isotropic colloids, where Poisson-like growth is always observed \cite{Yunker2013}. 
Our result suggests that the directionality of the interactions leads, always, to a self-affine interface.
The behavior for $\delta \gtrapprox \delta_\mathrm{min}$ is strongly affected by finite-size effects due to the proximity of the critical point.

We now characterize each transition in detail and estimate the
thresholds.

\subsection{Transition at $\delta_\mathrm{min}$} 
As explained before, when all patches are in the same hemisphere the
growth is eventually suppressed. Patches of colloids in the network are
typically pointing towards the substrate and are thus inaccessible for incoming
colloids ballistically approaching the substrate. Considering only
the geometrical effect, one expects
$\delta_\mathrm{min}=1/2-\theta/\pi=1/3$ (in units of $\pi\text{ rad}$),
where the first term refers to the equator and the second to the
interaction range. 

\begin{figure}[t]
  \begin{center}
   \includegraphics[width=\columnwidth]{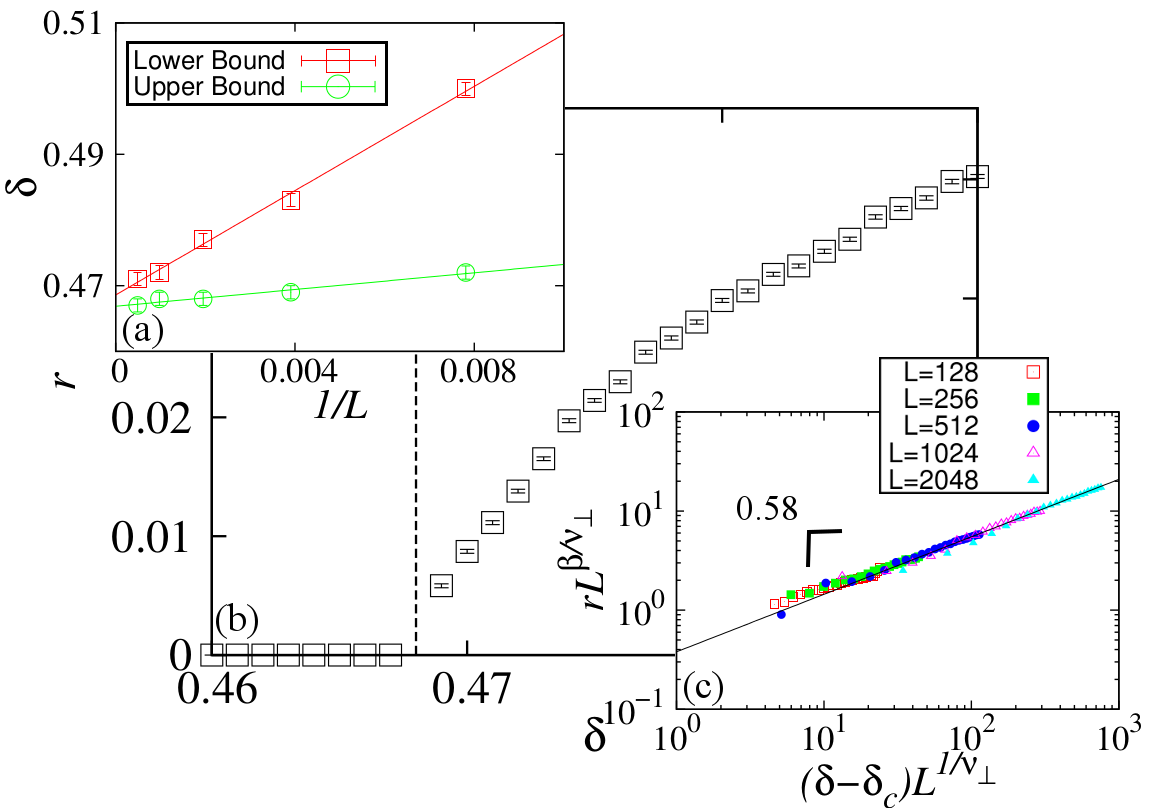}
  \end{center}
\caption{(color online) (a) Lower and upper bounds for $\delta_\mathrm{min}$ as a
function of $1/L$, for $L=\{128,256,512,1024,2048\}$ and
$\{1600,800,400,200,100\}$ samples. (b) Dependence of the growth rate
($r$) on the opening angle ($\delta$) for $L=2048$ averaged over $10^2$
samples. (c) Finite-size scaling for the growth rate, where
$\beta=0.58$, $\nu_\perp=0.73$, and $\delta_c=0.468$, consistent with
the Directed Percolation universality class. The system sizes and number
of samples are the same as in (a).  \label{fig.saturation_second} }
\end{figure}
To estimate the threshold, we performed simulations for
different values of $\delta$. For each value we ran several samples and
attempt the adsorption or binding of $2048L$ colloids. We considered
that growth is suppressed for runs where no attempt is successful
after $64L$ consecutive attempts. Due to strong finite-size effects, as
well known for absorbing-phase
transitions~\cite{Jensen1990a,Alencar1997}, growth is suppressed close to
$\delta_\mathrm{min}$ for a fraction of the
samples. We have used two different estimators for the threshold. The
lower bound ($\delta_\mathrm{min}^\mathrm{lb}$) is defined as the
highest $\delta$ for which growth was suppressed in every sample. The
upper bound ($\delta_\mathrm{min}^\mathrm{ub}$) is defined as the lowest
$\delta$ for which growth was never suppressed.
Figure~\ref{fig.saturation_second}(a) shows the value of both estimators
as a function of $1/L$. The linear extrapolation for the thermodynamic
limit ($L\rightarrow\infty$) gives
$\delta_\mathrm{min}^\mathrm{lb}=0.467\pm0.001$ and
$\delta_\mathrm{min}^\mathrm{ub}=0.469\pm0.002$. Combining these we
obtain $\delta_\mathrm{min}= 0.468\pm0.001$. Due to collective
effects during growth, the threshold is higher than that predicted by
a purely geometric argument for a single colloid.

A question of practical interest is how fast does the network grow.
For stochastic growth models, this can be assessed from the growth rate
($r$), defined as the fraction of successful adsorption/binding
attempts~\cite{AaraoReis2002}.  Figure~\ref{fig.saturation_second}(b)
shows the dependence of $r$ on $\delta$. $r$ grows
continuously from zero for $\delta\geq\delta_\mathrm{min}$, meaning that
the larger the angle the faster the network grows in mass. To identify
the universality class of the absorbing-phase transition at
$\delta_\mathrm{min}$ we use $r$ as the order parameter, which is zero
in the absorbing phase and non-zero in the active one.
Figure~\ref{fig.saturation_second}(c) depicts the finite-size scaling of the
order parameter. A data collapse is obtained over almost three
decades with the exponents of the Directed Percolation (DP)
universality class in two
dimensions~\cite{Henkel2008b,Lubeck2003,Odor2004}.

\begin{figure}[t]
  \begin{center}
   \includegraphics[width=\columnwidth]{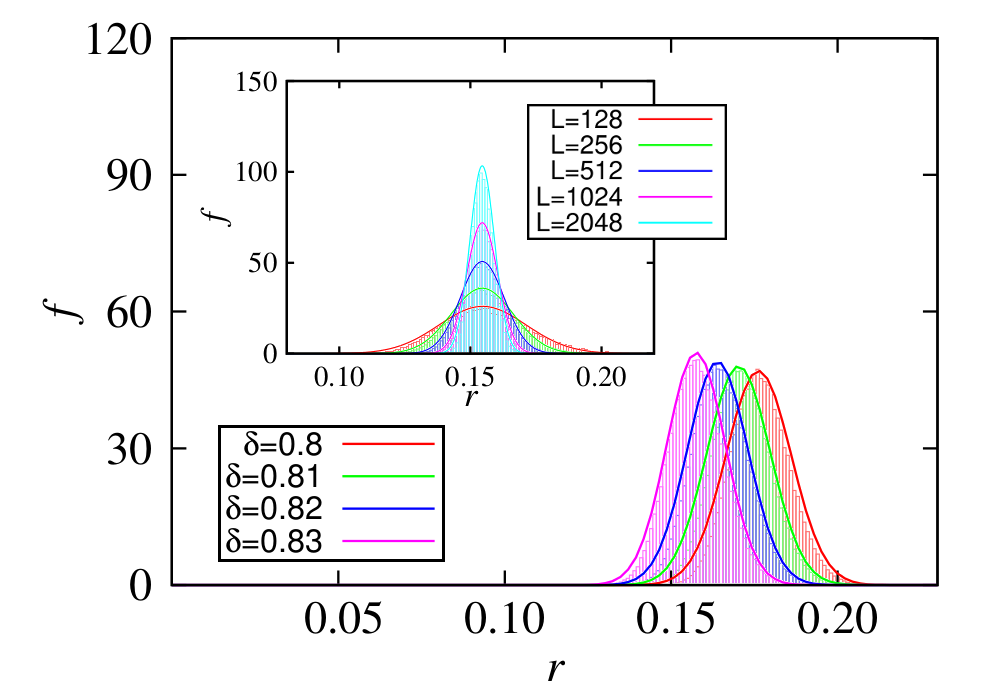} \\
  \end{center}
\caption{(color online) Main plot: Histogram of the growth rate for four values of
opening angle close to $\delta_\mathrm{max}$, namely,
$\delta=\{0.83,0.82,0.81,0.8\}$, and $L=512$, averaged over $10^5$
samples. Inset: Histogram of the growth rate at
$\delta_\mathrm{max}=\frac{5}{6}$ for $L=\{128,256,512,1024,2048\}$,
averaged over $\{16,8,4,2,1\}\times10^5$ samples.
\label{fig.saturation_first}}
\end{figure}
\subsection{Transition at $\delta_\mathrm{max}$} 
The second transition occurs when the distance between the two adjustable patches
is such that binding in only possible with one of them. We then expect
$\delta_\mathrm{max}=5/6\approx0.83$, a value that we have confirmed numerically by, 
as in the continuous case, performing a finite-size study of the transition point (as shown in the inset of 
Fig.~\ref{fig.saturation_first}). To describe this transition we also use $r$
as the order parameter.  The main plot of
Fig.~\ref{fig.saturation_first} is the histogram of $r$ for
different values of $\delta$. While in the absorbing phase
($\delta>\delta_\mathrm{max}$) $r=0$ (not shown), for
$\delta<\delta_\mathrm{max}$ $r$ is guassianly distributed
with a non-zero mean, which converges to $0.155\pm0.001$ at the
threshold. In the inset, we show the histogram of $r$ at
$\delta_\mathrm{max}$, for different system sizes. The larger the
system the sharper the distribution. The position of the peak does not
show significant size effects, and hence a jump is expected in the
thermodynamic limit. Thus, by contrast to the first transition, at
$\delta_\mathrm{max}$ the transition is discontinuous and the growth
rate jumps at the threshold. Note that, while in the vicinity of
$\delta_\mathrm{min}$, the growth rate vanishes with the substrate size,
at $\delta_\mathrm{max}$ it does not depend (significantly) on it.

\begin{figure}[t]
  \begin{center}
   \includegraphics[width=0.5\columnwidth]{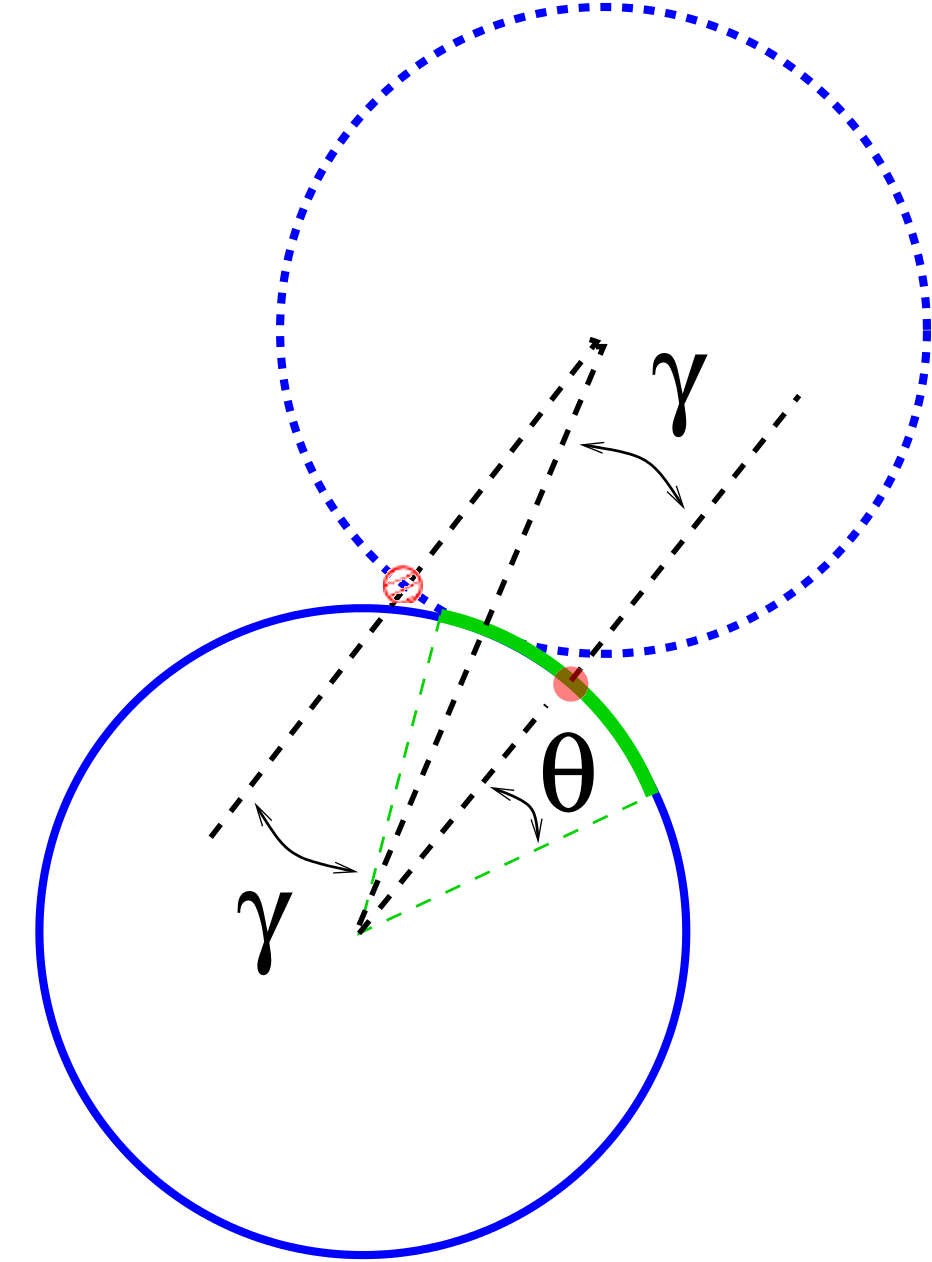} \\
   \end{center}
   \caption{(color online) Schematic
representation of the flexibility mechanism between bonds: the patch attempting binding will be at a
position defined by the angle $\gamma$ generated by a
Gaussian distribution centered at zero, with dispersion $F\theta$,
truncated at $F\theta$. For both colloids, the orientation of the bond is shifted from the ideal 
one by a randomly generated angle $\gamma$.
\label{fig.model_flexibility}}
 \end{figure}

\subsection{Flexibility} 
So far we have considered optimal bonds. This implies that the position of the incoming colloid is adjusted such 
that the center of the colloids and of their patches is aligned. However, even for chemical bonds, there is some 
flexibility around the optimal orientation \cite{Loweth1999,Geerts2010}. To model non-optimal bonds we take advantage of 
the stochastic nature of our model where the relative position and orientation of the colloids after 
collision may be adjusted. As in the optimal case, since the position and orientation of the network 
colloid are fixed, only the incoming colloid is adjusted. At a binding event, the flexibility for 
both rotation and translation of the colloid are represented by an angle $\gamma$ (see Fig.~\ref{fig.model_flexibility}). 
Inspired by previous models of patchy and DNA-mediated bonds \cite{Largo2007,Wilber2007},
the value of $\gamma$ is drawn randomly from a Gaussian distribution of zero mean and dispersion 
$F\theta$, where $F$ is the flexibility. Since the patch-patch interaction is short ranged, we truncate the 
distribution at $\mathrm{max}\{F\theta,\theta\}$. The sense of rotation of $\gamma$ is always from the center 
of the patch to the point of collision, as illustrated in Fig.~\ref{fig.model_flexibility}.

\begin{figure}[t]
  \begin{center}
   \includegraphics[width=\columnwidth]{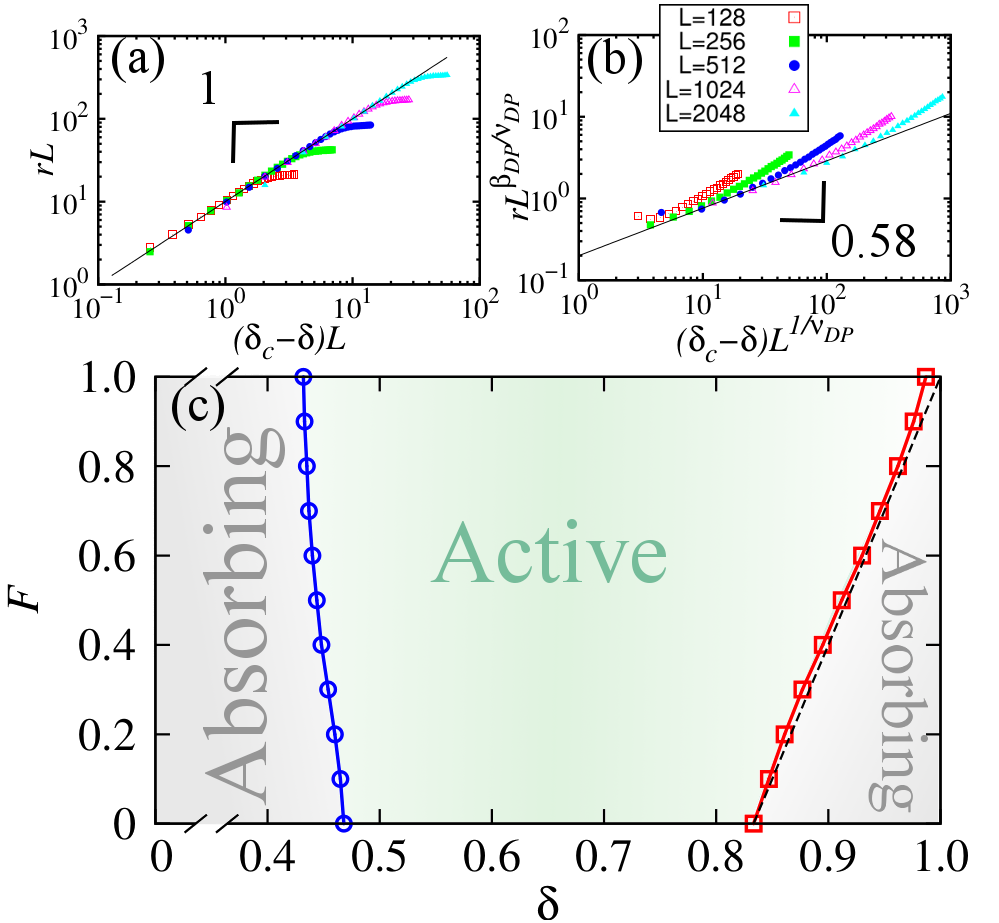} \\
  \end{center}
 \caption{(color online) Finite-size scaling for the growth rate rescaled 
 by the exponents of the absorbing transition for (a) $F=0.1$ 
 with linear rescaling and (b) $F=0.5$ with Directed Percolation rescaling.
 Results are for $L=\{128,256,512,1024,2048\}$ and averaged
over $\{1600,800,400,200,100\}$ samples. (c) Phase diagram in the
two-parameter space: flexibility ($F$) and opening angle ($\delta$). The
(blue-)solid curve in the left-hand side corresponds to the lower
threshold ($\delta_\mathrm{min}$) and the (red-)solid curve in the
right-hand side to the higher one ($\delta_\mathrm{max}$). The data
points are extrapolations for the thermodynamic limit from the behavior
of the size dependence of the thresholds. The (black-)dashed curve is
the theoretical prediction for $\delta_\mathrm{max}$.
\label{fig.flexibility}}
\end{figure}

We performed simulations for different values of $F$ and
$\delta$. For the first transition, the value of $\delta_\mathrm{min}$
slightly decreases with $F$ (see (blue-)solid curve on the
left-hand side of Fig.~\ref{fig.flexibility}(c)). Yet, for the range of
flexibilities considered here the transition is always continuous and
in the DP universality class.

For the second transition, the value of $\delta_\mathrm{max}$ increases
with $F$ ((red-)solid curve on the right-hand side of
Fig.~\ref{fig.flexibility}(c)). Hence, the range of $\delta$ for
which growth is sustained increased with $F$. For $F=0$ we have
shown that the threshold corresponds to the opening angle when two
colloids bound to the adjustable patches touch. likewise,
we can estimate the threshold $\delta_\mathrm{max}(F)$ for
general $F$,
\begin{equation}\label{eq.threshold.F}
\delta_\mathrm{max}(F)=\delta_\mathrm{max}(0)+\frac{F\theta}{\pi} \ \ ,
\end{equation}
where $\delta_\mathrm{max}(0)=5/6$ (as discussed before) and the second
term corresponds to the maximum value of $\gamma$. As shown in
Fig.~\ref{fig.flexibility}(c) ((black-)dashed curve), the threshold
values differ by less than $2\%$ from the
theoretical prediction. In fact, the difference between the numerical
and theoretical values vanishes with decreasing $F$ and
Eq.~(\ref{eq.threshold.F}) is exact for $F=0$.

The nature of the transition also changes with the $F$.
Figure~\ref{fig.flexibility}(a)~and~(b) depict the data collapses for the
order parameter at $F=0.1$ and $F=0.5$. At $F=0.1$ the transition is
still discontinuous and data collapse is obtained with trivial
exponents (see Fig.~\ref{fig.flexibility}(a)). By contrast, at $F=0.5$
the transition is continuous in the DP universality class, as
evident from Fig.~\ref{fig.flexibility}(b). We then expect the
nature of the transition to change at a tricritical flexibility (between
$0.1$ and $0.5$), in the tricritical Directed
Percolation universality class~\cite{Lubeck2006,Grassberger2006}.
The limit $F\rightarrow\infty$ corresponds to a uniform distribution of the bonds over 
the interaction range. In this limit, the active region in the diagram of Fig.~\ref{fig.flexibility}(c) is maximal.

Figure~\ref{fig.flexibility}(a) and (b) depict two different data collapses, for the order parameter at 
$F=0.1$ and $F=0.5$, with different rescaling of the vertical axis. 
At $F=0.1$ the transition is still discontinuous and data collapse is obtained with linear rescaling typical of a 
discontinuous transition (see Fig.~\ref{fig.flexibility}(a)). By contrast, at $F=0.5$ the transition is continuous, and data collapse 
is obtained by rescaling the vertical axis with the exponents of the DP universality class, as shown in Fig.~\ref{fig.flexibility}(b).

\section{Conclusion}\label{sec.conc}
We found that the adsorption of patchy
colloids on substrates depends strongly on the opening angle between
patches. The growth of a colloidal network from the substrate is only
sustained between minimum and maximum opening angles. Outside of
this active phase the system is trapped into one of two possible absorbing
phases where growth is suppressed at a finite thickness of the network.
The transitions into the two absorbing phases are quite different.
While the transition at the lower threshold is continuous in the
DP universality class, that at the higher
threshold is discontinuous. We provided an estimator of the higher
threshold which is exact in some limits. We also showed that the nature
of the transition is intimately related to the growth rate of the
network. For a continuous transition the growth rate vanishes in the
vicinity of the threshold, while for a discontinuous transition the
growth rate has a jump. This difference has obvious practical
implications on the feasibility of the predicted structures. We have shown that it is possible to effectively
control the interface roughness by varying the opening angle. We are also able to widen the active region of 
growth by increasing the flexibility of the bonds. 

The numerical results were obtained for a two dimensional system but our conclusions may be extended to three dimensions. 
However, recent experimental work on colloidal 
aggregation at the edge of an evaporating drop, may be described as a two dimensional system \cite{Yunker2013,Yunker2013b,Yang2014}. 
Such drops may provide a direct experimental realization of our model if patchy colloids with strong bonds are used.

Absorbing phase transitions are the focus of many
theoretical~\cite{Ziff1986,Marro1999,AaraoReis2002,Henkel2008b} and
recent experimental works~\cite{Takeuchi2014,Takeuchi2007,Shi2013},
including studies that successfully combine
both~\cite{Pine2005,Corte2008}. Special attention has been given to
nonequilibrium wetting transitions. Also there a change in the nature of
the transition is observed at a multicritical strength of the attraction to
the substrate. While mesoscopic models predict that the second continuous
transition typically falls into the Multiplicative Noise (MN1)
universality class~\cite{Hinrichsen2000b,Muoz2005}, they also identify DP 
transitions~\cite{Ginelli2003}. In fact, a crossover from MN1 to
DP is expected when varying the control parameter~\cite{Ginelli2003}.
However, the nonequilibrium wetting phenomenon typically
involves three scaling fields, for example, temperature,
chemical potential, and surface potential. By contrast, the tricritical
transition for patchy colloids presented here is driven by the colloid parameters, 
namely, the opening angle and flexibility. More importantly, 
the fluctuations in non-equilibrium wetting models are different from the
inherent in our model. Thus,
there are interesting possible follow ups. The identification
of the third scaling field, the effect of other fluctuations, and the study of the scaling at the
tricritical flexibility. 
More generally, if desorption and/or bulk thermal fluctuations are included, 
the transitions between the thin and thick adsorbed films may be related to non-equilibrium wetting phenomena \cite{DelosSantos2002,DelosSantos2003}.

\begin{acknowledgments} 
We acknowledge financial support from the
Portuguese Foundation for Science and Technology (FCT) under Contracts
nos. EXCL/FIS-NAN/0083/2012, PEst-OE/FIS/UI0618/2014, and IF/00255/2013. 
\end{acknowledgments}

\bibliography{PatchyColloids}

\begin{thebibliography}{73}%
\makeatletter
\providecommand \@ifxundefined [1]{%
 \@ifx{#1\undefined}
}%
\providecommand \@ifnum [1]{%
 \ifnum #1\expandafter \@firstoftwo
 \else \expandafter \@secondoftwo
 \fi
}%
\providecommand \@ifx [1]{%
 \ifx #1\expandafter \@firstoftwo
 \else \expandafter \@secondoftwo
 \fi
}%
\providecommand \natexlab [1]{#1}%
\providecommand \enquote  [1]{``#1''}%
\providecommand \bibnamefont  [1]{#1}%
\providecommand \bibfnamefont [1]{#1}%
\providecommand \citenamefont [1]{#1}%
\providecommand \href@noop [0]{\@secondoftwo}%
\providecommand \href [0]{\begingroup \@sanitize@url \@href}%
\providecommand \@href[1]{\@@startlink{#1}\@@href}%
\providecommand \@@href[1]{\endgroup#1\@@endlink}%
\providecommand \@sanitize@url [0]{\catcode `\\12\catcode `\$12\catcode
  `\&12\catcode `\#12\catcode `\^12\catcode `\_12\catcode `\%12\relax}%
\providecommand \@@startlink[1]{}%
\providecommand \@@endlink[0]{}%
\providecommand \url  [0]{\begingroup\@sanitize@url \@url }%
\providecommand \@url [1]{\endgroup\@href {#1}{\urlprefix }}%
\providecommand \urlprefix  [0]{URL }%
\providecommand \Eprint [0]{\href }%
\providecommand \doibase [0]{http://dx.doi.org/}%
\providecommand \selectlanguage [0]{\@gobble}%
\providecommand \bibinfo  [0]{\@secondoftwo}%
\providecommand \bibfield  [0]{\@secondoftwo}%
\providecommand \translation [1]{[#1]}%
\providecommand \BibitemOpen [0]{}%
\providecommand \bibitemStop [0]{}%
\providecommand \bibitemNoStop [0]{.\EOS\space}%
\providecommand \EOS [0]{\spacefactor3000\relax}%
\providecommand \BibitemShut  [1]{\csname bibitem#1\endcsname}%
\let\auto@bib@innerbib\@empty
%</preamble>
\bibitem [{\citenamefont {van Blaaderen}(2006)}]{Blaaderen2006}%
  \BibitemOpen
  \bibfield  {author} {\bibinfo {author} {\bibfnamefont {A.}~\bibnamefont {van
  Blaaderen}},\ }\href@noop {} {\bibfield  {journal} {\bibinfo  {journal}
  {Nature}\ }\textbf {\bibinfo {volume} {439}},\ \bibinfo {pages} {545}
  (\bibinfo {year} {2006})}\BibitemShut {NoStop}%
\bibitem [{\citenamefont {Bianchi}\ \emph {et~al.}(2011)\citenamefont
  {Bianchi}, \citenamefont {Blaak},\ and\ \citenamefont {Likos}}]{Bianchi2011}%
  \BibitemOpen
  \bibfield  {author} {\bibinfo {author} {\bibfnamefont {E.}~\bibnamefont
  {Bianchi}}, \bibinfo {author} {\bibfnamefont {R.}~\bibnamefont {Blaak}}, \
  and\ \bibinfo {author} {\bibfnamefont {C.~N.}\ \bibnamefont {Likos}},\
  }\href@noop {} {\bibfield  {journal} {\bibinfo  {journal} {Phys. Chem. Chem.
  Phys.}\ }\textbf {\bibinfo {volume} {13}},\ \bibinfo {pages} {6397} (\bibinfo
  {year} {2011})}\BibitemShut {NoStop}%
\bibitem [{\citenamefont {Pawar}\ and\ \citenamefont
  {Kretzschmar}(2010)}]{Pawar2010}%
  \BibitemOpen
  \bibfield  {author} {\bibinfo {author} {\bibfnamefont {A.~B.}\ \bibnamefont
  {Pawar}}\ and\ \bibinfo {author} {\bibfnamefont {I.}~\bibnamefont
  {Kretzschmar}},\ }\href@noop {} {\bibfield  {journal} {\bibinfo  {journal}
  {Macromol. Rapid Commun.}\ }\textbf {\bibinfo {volume} {31}},\ \bibinfo
  {pages} {150} (\bibinfo {year} {2010})}\BibitemShut {NoStop}%
\bibitem [{\citenamefont {Kretzschmar}\ and\ \citenamefont
  {Song}(2011)}]{Kretzschmar2011}%
  \BibitemOpen
  \bibfield  {author} {\bibinfo {author} {\bibfnamefont {I.}~\bibnamefont
  {Kretzschmar}}\ and\ \bibinfo {author} {\bibfnamefont {J.~H.}\ \bibnamefont
  {Song}},\ }\href@noop {} {\bibfield  {journal} {\bibinfo  {journal} {Curr.
  Opin. Coll. Interf. Sci.}\ }\textbf {\bibinfo {volume} {16}},\ \bibinfo
  {pages} {84} (\bibinfo {year} {2011})}\BibitemShut {NoStop}%
\bibitem [{\citenamefont {Grzelczak}\ \emph {et~al.}(2010)\citenamefont
  {Grzelczak}, \citenamefont {Vermant}, \citenamefont {Furst},\ and\
  \citenamefont {Liz-Marz\'{a}n}}]{Grzelczak2010}%
  \BibitemOpen
  \bibfield  {author} {\bibinfo {author} {\bibfnamefont {M.}~\bibnamefont
  {Grzelczak}}, \bibinfo {author} {\bibfnamefont {J.}~\bibnamefont {Vermant}},
  \bibinfo {author} {\bibfnamefont {E.~M.}\ \bibnamefont {Furst}}, \ and\
  \bibinfo {author} {\bibfnamefont {L.~M.}\ \bibnamefont {Liz-Marz\'{a}n}},\
  }\href@noop {} {\bibfield  {journal} {\bibinfo  {journal} {ACS nano}\
  }\textbf {\bibinfo {volume} {4}},\ \bibinfo {pages} {3591} (\bibinfo {year}
  {2010})}\BibitemShut {NoStop}%
\bibitem [{\citenamefont {Wilner}\ and\ \citenamefont
  {Willner}(2012)}]{Wilner2012}%
  \BibitemOpen
  \bibfield  {author} {\bibinfo {author} {\bibfnamefont {O.~I.}\ \bibnamefont
  {Wilner}}\ and\ \bibinfo {author} {\bibfnamefont {I.}~\bibnamefont
  {Willner}},\ }\href@noop {} {\bibfield  {journal} {\bibinfo  {journal} {Chem.
  Rev.}\ }\textbf {\bibinfo {volume} {112}},\ \bibinfo {pages} {2528} (\bibinfo
  {year} {2012})}\BibitemShut {NoStop}%
\bibitem [{\citenamefont {Yi}\ \emph {et~al.}(2013)\citenamefont {Yi},
  \citenamefont {Pine},\ and\ \citenamefont {Sacanna}}]{Yi2013}%
  \BibitemOpen
  \bibfield  {author} {\bibinfo {author} {\bibfnamefont {G.-R.}\ \bibnamefont
  {Yi}}, \bibinfo {author} {\bibfnamefont {D.~J.}\ \bibnamefont {Pine}}, \ and\
  \bibinfo {author} {\bibfnamefont {S.}~\bibnamefont {Sacanna}},\ }\href@noop
  {} {\bibfield  {journal} {\bibinfo  {journal} {J. Phys.: Condens. Matter}\
  }\textbf {\bibinfo {volume} {25}},\ \bibinfo {pages} {193101} (\bibinfo
  {year} {2013})}\BibitemShut {NoStop}%
\bibitem [{\citenamefont {Wang}\ \emph {et~al.}(2012)\citenamefont {Wang},
  \citenamefont {Breed}, \citenamefont {Manoharan}, \citenamefont {Feng},
  \citenamefont {Hollingsworth}, \citenamefont {Weck},\ and\ \citenamefont
  {Pine}}]{Wang2012}%
  \BibitemOpen
  \bibfield  {author} {\bibinfo {author} {\bibfnamefont {Y.}~\bibnamefont
  {Wang}}, \bibinfo {author} {\bibfnamefont {D.~R.}\ \bibnamefont {Breed}},
  \bibinfo {author} {\bibfnamefont {V.~N.}\ \bibnamefont {Manoharan}}, \bibinfo
  {author} {\bibfnamefont {L.}~\bibnamefont {Feng}}, \bibinfo {author}
  {\bibfnamefont {A.~D.}\ \bibnamefont {Hollingsworth}}, \bibinfo {author}
  {\bibfnamefont {M.}~\bibnamefont {Weck}}, \ and\ \bibinfo {author}
  {\bibfnamefont {D.~J.}\ \bibnamefont {Pine}},\ }\href@noop {} {\bibfield
  {journal} {\bibinfo  {journal} {Nature}\ }\textbf {\bibinfo {volume} {491}},\
  \bibinfo {pages} {51} (\bibinfo {year} {2012})}\BibitemShut {NoStop}%
\bibitem [{\citenamefont {Bianchi}\ \emph {et~al.}(2006)\citenamefont
  {Bianchi}, \citenamefont {Largo}, \citenamefont {Tartaglia}, \citenamefont
  {Zaccarelli},\ and\ \citenamefont {Sciortino}}]{Bianchi2006}%
  \BibitemOpen
  \bibfield  {author} {\bibinfo {author} {\bibfnamefont {E.}~\bibnamefont
  {Bianchi}}, \bibinfo {author} {\bibfnamefont {J.}~\bibnamefont {Largo}},
  \bibinfo {author} {\bibfnamefont {P.}~\bibnamefont {Tartaglia}}, \bibinfo
  {author} {\bibfnamefont {E.}~\bibnamefont {Zaccarelli}}, \ and\ \bibinfo
  {author} {\bibfnamefont {F.}~\bibnamefont {Sciortino}},\ }\href@noop {}
  {\bibfield  {journal} {\bibinfo  {journal} {Phys. Rev. Lett.}\ }\textbf
  {\bibinfo {volume} {97}},\ \bibinfo {pages} {168301} (\bibinfo {year}
  {2006})}\BibitemShut {NoStop}%
\bibitem [{\citenamefont {Russo}\ \emph {et~al.}(2009)\citenamefont {Russo},
  \citenamefont {Tartaglia},\ and\ \citenamefont {Sciortino}}]{Russo2009}%
  \BibitemOpen
  \bibfield  {author} {\bibinfo {author} {\bibfnamefont {J.}~\bibnamefont
  {Russo}}, \bibinfo {author} {\bibfnamefont {P.}~\bibnamefont {Tartaglia}}, \
  and\ \bibinfo {author} {\bibfnamefont {F.}~\bibnamefont {Sciortino}},\
  }\href@noop {} {\bibfield  {journal} {\bibinfo  {journal} {J. Chem. Phys.}\
  }\textbf {\bibinfo {volume} {131}},\ \bibinfo {pages} {014504} (\bibinfo
  {year} {2009})}\BibitemShut {NoStop}%
\bibitem [{\citenamefont {Tavares}\ \emph
  {et~al.}(2009{\natexlab{a}})\citenamefont {Tavares}, \citenamefont
  {Teixeira},\ and\ \citenamefont {{Telo da Gama}}}]{Tavares2009}%
  \BibitemOpen
  \bibfield  {author} {\bibinfo {author} {\bibfnamefont {J.~M.}\ \bibnamefont
  {Tavares}}, \bibinfo {author} {\bibfnamefont {P.~I.~C.}\ \bibnamefont
  {Teixeira}}, \ and\ \bibinfo {author} {\bibfnamefont {M.~M.}\ \bibnamefont
  {{Telo da Gama}}},\ }\href@noop {} {\bibfield  {journal} {\bibinfo  {journal}
  {Mol. Phys.}\ }\textbf {\bibinfo {volume} {107}},\ \bibinfo {pages} {453}
  (\bibinfo {year} {2009}{\natexlab{a}})}\BibitemShut {NoStop}%
\bibitem [{\citenamefont {Tavares}\ \emph
  {et~al.}(2009{\natexlab{b}})\citenamefont {Tavares}, \citenamefont
  {Teixeira},\ and\ \citenamefont {{Telo da Gama}}}]{Tavares2009a}%
  \BibitemOpen
  \bibfield  {author} {\bibinfo {author} {\bibfnamefont {J.~M.}\ \bibnamefont
  {Tavares}}, \bibinfo {author} {\bibfnamefont {P.~I.~C.}\ \bibnamefont
  {Teixeira}}, \ and\ \bibinfo {author} {\bibfnamefont {M.~M.}\ \bibnamefont
  {{Telo da Gama}}},\ }\href@noop {} {\bibfield  {journal} {\bibinfo  {journal}
  {Phys. Rev. E}\ }\textbf {\bibinfo {volume} {80}},\ \bibinfo {pages} {021506}
  (\bibinfo {year} {2009}{\natexlab{b}})}\BibitemShut {NoStop}%
\bibitem [{\citenamefont {de~las Heras}\ \emph {et~al.}(2011)\citenamefont
  {de~las Heras}, \citenamefont {Tavares},\ and\ \citenamefont {{Telo da
  Gama}}}]{DelasHeras2011}%
  \BibitemOpen
  \bibfield  {author} {\bibinfo {author} {\bibfnamefont {D.}~\bibnamefont
  {de~las Heras}}, \bibinfo {author} {\bibfnamefont {J.~M.}\ \bibnamefont
  {Tavares}}, \ and\ \bibinfo {author} {\bibfnamefont {M.~M.}\ \bibnamefont
  {{Telo da Gama}}},\ }\href@noop {} {\bibfield  {journal} {\bibinfo  {journal}
  {Soft Matt.}\ }\textbf {\bibinfo {volume} {7}},\ \bibinfo {pages} {5615}
  (\bibinfo {year} {2011})}\BibitemShut {NoStop}%
\bibitem [{\citenamefont {Hu}\ \emph {et~al.}(2012)\citenamefont {Hu},
  \citenamefont {Zhou}, \citenamefont {Sun}, \citenamefont {Fang},\ and\
  \citenamefont {Wu}}]{Hu2012}%
  \BibitemOpen
  \bibfield  {author} {\bibinfo {author} {\bibfnamefont {J.}~\bibnamefont
  {Hu}}, \bibinfo {author} {\bibfnamefont {S.}~\bibnamefont {Zhou}}, \bibinfo
  {author} {\bibfnamefont {Y.}~\bibnamefont {Sun}}, \bibinfo {author}
  {\bibfnamefont {X.}~\bibnamefont {Fang}}, \ and\ \bibinfo {author}
  {\bibfnamefont {L.}~\bibnamefont {Wu}},\ }\href@noop {} {\bibfield  {journal}
  {\bibinfo  {journal} {Chem. Soc. Rev.}\ }\textbf {\bibinfo {volume} {41}},\
  \bibinfo {pages} {4356} (\bibinfo {year} {2012})}\BibitemShut {NoStop}%
\bibitem [{\citenamefont {Angioletti-Uberti}\ \emph {et~al.}(2012)\citenamefont
  {Angioletti-Uberti}, \citenamefont {Mognetti},\ and\ \citenamefont
  {Frenkel}}]{Angioletti-Uberti2012}%
  \BibitemOpen
  \bibfield  {author} {\bibinfo {author} {\bibfnamefont {S.}~\bibnamefont
  {Angioletti-Uberti}}, \bibinfo {author} {\bibfnamefont {B.~M.}\ \bibnamefont
  {Mognetti}}, \ and\ \bibinfo {author} {\bibfnamefont {D.}~\bibnamefont
  {Frenkel}},\ }\href@noop {} {\bibfield  {journal} {\bibinfo  {journal}
  {Nature Mater.}\ }\textbf {\bibinfo {volume} {11}},\ \bibinfo {pages} {518}
  (\bibinfo {year} {2012})}\BibitemShut {NoStop}%
\bibitem [{\citenamefont {Kumar}\ \emph {et~al.}(2013)\citenamefont {Kumar},
  \citenamefont {Park}, \citenamefont {Tu},\ and\ \citenamefont
  {Lee}}]{Kumar2013}%
  \BibitemOpen
  \bibfield  {author} {\bibinfo {author} {\bibfnamefont {A.}~\bibnamefont
  {Kumar}}, \bibinfo {author} {\bibfnamefont {B.~J.}\ \bibnamefont {Park}},
  \bibinfo {author} {\bibfnamefont {F.}~\bibnamefont {Tu}}, \ and\ \bibinfo
  {author} {\bibfnamefont {D.}~\bibnamefont {Lee}},\ }\href@noop {} {\bibfield
  {journal} {\bibinfo  {journal} {Soft Matt.}\ }\textbf {\bibinfo {volume}
  {9}},\ \bibinfo {pages} {6604} (\bibinfo {year} {2013})}\BibitemShut
  {NoStop}%
\bibitem [{\citenamefont {Russo}\ \emph {et~al.}(2011)\citenamefont {Russo},
  \citenamefont {Tavares}, \citenamefont {Teixeira}, \citenamefont {{Telo da
  Gama}},\ and\ \citenamefont {Sciortino}}]{Russo2011a}%
  \BibitemOpen
  \bibfield  {author} {\bibinfo {author} {\bibfnamefont {J.}~\bibnamefont
  {Russo}}, \bibinfo {author} {\bibfnamefont {J.~M.}\ \bibnamefont {Tavares}},
  \bibinfo {author} {\bibfnamefont {P.~I.~C.}\ \bibnamefont {Teixeira}},
  \bibinfo {author} {\bibfnamefont {M.~M.}\ \bibnamefont {{Telo da Gama}}}, \
  and\ \bibinfo {author} {\bibfnamefont {F.}~\bibnamefont {Sciortino}},\
  }\href@noop {} {\bibfield  {journal} {\bibinfo  {journal} {Phys. Rev. Lett.}\
  }\textbf {\bibinfo {volume} {106}},\ \bibinfo {pages} {085703} (\bibinfo
  {year} {2011})}\BibitemShut {NoStop}%
\bibitem [{\citenamefont {{de Las Heras}}\ \emph {et~al.}(2012)\citenamefont
  {{de Las Heras}}, \citenamefont {Tavares},\ and\ \citenamefont {{Telo da
  Gama}}}]{DelasHeras2012}%
  \BibitemOpen
  \bibfield  {author} {\bibinfo {author} {\bibfnamefont {D.}~\bibnamefont {{de
  Las Heras}}}, \bibinfo {author} {\bibfnamefont {J.~M.}\ \bibnamefont
  {Tavares}}, \ and\ \bibinfo {author} {\bibfnamefont {M.~M.}\ \bibnamefont
  {{Telo da Gama}}},\ }\href@noop {} {\bibfield  {journal} {\bibinfo  {journal}
  {Soft Matt.}\ }\textbf {\bibinfo {volume} {8}},\ \bibinfo {pages} {1785}
  (\bibinfo {year} {2012})}\BibitemShut {NoStop}%
\bibitem [{\citenamefont {Reinhardt}\ \emph {et~al.}(2013)\citenamefont
  {Reinhardt}, \citenamefont {Romano},\ and\ \citenamefont
  {Doye}}]{Reinhardt2013}%
  \BibitemOpen
  \bibfield  {author} {\bibinfo {author} {\bibfnamefont {A.}~\bibnamefont
  {Reinhardt}}, \bibinfo {author} {\bibfnamefont {F.}~\bibnamefont {Romano}}, \
  and\ \bibinfo {author} {\bibfnamefont {J.~P.~K.}\ \bibnamefont {Doye}},\
  }\href@noop {} {\bibfield  {journal} {\bibinfo  {journal} {Phys. Rev. Lett.}\
  }\textbf {\bibinfo {volume} {110}},\ \bibinfo {pages} {255503} (\bibinfo
  {year} {2013})}\BibitemShut {NoStop}%
\bibitem [{\citenamefont {Smallenburg}\ and\ \citenamefont
  {Sciortino}(2013)}]{Smallenburg2013}%
  \BibitemOpen
  \bibfield  {author} {\bibinfo {author} {\bibfnamefont {F.}~\bibnamefont
  {Smallenburg}}\ and\ \bibinfo {author} {\bibfnamefont {F.}~\bibnamefont
  {Sciortino}},\ }\href@noop {} {\bibfield  {journal} {\bibinfo  {journal}
  {Nature Phys.}\ }\textbf {\bibinfo {volume} {9}},\ \bibinfo {pages} {554}
  (\bibinfo {year} {2013})}\BibitemShut {NoStop}%
\bibitem [{\citenamefont {Matthews}\ and\ \citenamefont
  {Likos}(2012)}]{Matthews2012}%
  \BibitemOpen
  \bibfield  {author} {\bibinfo {author} {\bibfnamefont {R.}~\bibnamefont
  {Matthews}}\ and\ \bibinfo {author} {\bibfnamefont {C.~N.}\ \bibnamefont
  {Likos}},\ }\href@noop {} {\bibfield  {journal} {\bibinfo  {journal} {Phys.
  Rev. Lett.}\ }\textbf {\bibinfo {volume} {109}},\ \bibinfo {pages} {178302}
  (\bibinfo {year} {2012})}\BibitemShut {NoStop}%
\bibitem [{\citenamefont {Smallenburg}\ \emph {et~al.}(2013)\citenamefont
  {Smallenburg}, \citenamefont {Leibler},\ and\ \citenamefont
  {Sciortino}}]{Smallenburg2013a}%
  \BibitemOpen
  \bibfield  {author} {\bibinfo {author} {\bibfnamefont {F.}~\bibnamefont
  {Smallenburg}}, \bibinfo {author} {\bibfnamefont {L.}~\bibnamefont
  {Leibler}}, \ and\ \bibinfo {author} {\bibfnamefont {F.}~\bibnamefont
  {Sciortino}},\ }\href@noop {} {\bibfield  {journal} {\bibinfo  {journal}
  {Phys. Rev. Lett.}\ }\textbf {\bibinfo {volume} {111}},\ \bibinfo {pages}
  {188002} (\bibinfo {year} {2013})}\BibitemShut {NoStop}%
\bibitem [{\citenamefont {Coluzza}\ \emph {et~al.}(2013)\citenamefont
  {Coluzza}, \citenamefont {van Oostrum}, \citenamefont {Capone}, \citenamefont
  {Reimhult},\ and\ \citenamefont {Dellago}}]{Coluzza2013}%
  \BibitemOpen
  \bibfield  {author} {\bibinfo {author} {\bibfnamefont {I.}~\bibnamefont
  {Coluzza}}, \bibinfo {author} {\bibfnamefont {P.~D.~J.}\ \bibnamefont {van
  Oostrum}}, \bibinfo {author} {\bibfnamefont {B.}~\bibnamefont {Capone}},
  \bibinfo {author} {\bibfnamefont {E.}~\bibnamefont {Reimhult}}, \ and\
  \bibinfo {author} {\bibfnamefont {C.}~\bibnamefont {Dellago}},\ }\href@noop
  {} {\bibfield  {journal} {\bibinfo  {journal} {Phys. Rev. Lett.}\ }\textbf
  {\bibinfo {volume} {110}},\ \bibinfo {pages} {075501} (\bibinfo {year}
  {2013})}\BibitemShut {NoStop}%
\bibitem [{\citenamefont {Sciortino}\ \emph {et~al.}(2009)\citenamefont
  {Sciortino}, \citenamefont {{De Michele}}, \citenamefont {Corezzi},
  \citenamefont {Russo}, \citenamefont {Zaccarelli},\ and\ \citenamefont
  {Tartaglia}}]{Sciortino2009}%
  \BibitemOpen
  \bibfield  {author} {\bibinfo {author} {\bibfnamefont {F.}~\bibnamefont
  {Sciortino}}, \bibinfo {author} {\bibfnamefont {C.}~\bibnamefont {{De
  Michele}}}, \bibinfo {author} {\bibfnamefont {S.}~\bibnamefont {Corezzi}},
  \bibinfo {author} {\bibfnamefont {J.}~\bibnamefont {Russo}}, \bibinfo
  {author} {\bibfnamefont {E.}~\bibnamefont {Zaccarelli}}, \ and\ \bibinfo
  {author} {\bibfnamefont {P.}~\bibnamefont {Tartaglia}},\ }\href@noop {}
  {\bibfield  {journal} {\bibinfo  {journal} {Soft Matter}\ }\textbf {\bibinfo
  {volume} {5}},\ \bibinfo {pages} {2571} (\bibinfo {year} {2009})}\BibitemShut
  {NoStop}%
\bibitem [{\citenamefont {Corezzi}\ \emph {et~al.}(2009)\citenamefont
  {Corezzi}, \citenamefont {{De Michele}}, \citenamefont {Zaccarelli},
  \citenamefont {Tartaglia},\ and\ \citenamefont {Sciortino}}]{Corezzi2009}%
  \BibitemOpen
  \bibfield  {author} {\bibinfo {author} {\bibfnamefont {S.}~\bibnamefont
  {Corezzi}}, \bibinfo {author} {\bibfnamefont {C.}~\bibnamefont {{De
  Michele}}}, \bibinfo {author} {\bibfnamefont {E.}~\bibnamefont {Zaccarelli}},
  \bibinfo {author} {\bibfnamefont {P.}~\bibnamefont {Tartaglia}}, \ and\
  \bibinfo {author} {\bibfnamefont {F.}~\bibnamefont {Sciortino}},\ }\href@noop
  {} {\bibfield  {journal} {\bibinfo  {journal} {J. Phys. Chem. B}\ }\textbf
  {\bibinfo {volume} {113}},\ \bibinfo {pages} {1233} (\bibinfo {year}
  {2009})}\BibitemShut {NoStop}%
\bibitem [{\citenamefont {Corezzi}\ \emph {et~al.}(2012)\citenamefont
  {Corezzi}, \citenamefont {Fioretto},\ and\ \citenamefont
  {Sciortino}}]{Corezzi2012}%
  \BibitemOpen
  \bibfield  {author} {\bibinfo {author} {\bibfnamefont {S.}~\bibnamefont
  {Corezzi}}, \bibinfo {author} {\bibfnamefont {D.}~\bibnamefont {Fioretto}}, \
  and\ \bibinfo {author} {\bibfnamefont {F.}~\bibnamefont {Sciortino}},\
  }\href@noop {} {\bibfield  {journal} {\bibinfo  {journal} {Soft Matt.}\
  }\textbf {\bibinfo {volume} {8}},\ \bibinfo {pages} {11207} (\bibinfo {year}
  {2012})}\BibitemShut {NoStop}%
\bibitem [{\citenamefont {Vasilyev}\ \emph {et~al.}(2013)\citenamefont
  {Vasilyev}, \citenamefont {Klumov},\ and\ \citenamefont
  {Tkachenko}}]{Vasilyev2013}%
  \BibitemOpen
  \bibfield  {author} {\bibinfo {author} {\bibfnamefont {O.~A.}\ \bibnamefont
  {Vasilyev}}, \bibinfo {author} {\bibfnamefont {B.~A.}\ \bibnamefont
  {Klumov}}, \ and\ \bibinfo {author} {\bibfnamefont {A.~V.}\ \bibnamefont
  {Tkachenko}},\ }\href@noop {} {\bibfield  {journal} {\bibinfo  {journal}
  {Phys. Rev. E}\ }\textbf {\bibinfo {volume} {88}},\ \bibinfo {pages} {012302}
  (\bibinfo {year} {2013})}\BibitemShut {NoStop}%
\bibitem [{\citenamefont {Gnan}\ \emph {et~al.}(2012)\citenamefont {Gnan},
  \citenamefont {{de Las Heras}}, \citenamefont {Tavares}, \citenamefont {{Telo
  da Gama}},\ and\ \citenamefont {Sciortino}}]{Gnan2012}%
  \BibitemOpen
  \bibfield  {author} {\bibinfo {author} {\bibfnamefont {N.}~\bibnamefont
  {Gnan}}, \bibinfo {author} {\bibfnamefont {D.}~\bibnamefont {{de Las
  Heras}}}, \bibinfo {author} {\bibfnamefont {J.~M.}\ \bibnamefont {Tavares}},
  \bibinfo {author} {\bibfnamefont {M.~M.}\ \bibnamefont {{Telo da Gama}}}, \
  and\ \bibinfo {author} {\bibfnamefont {F.}~\bibnamefont {Sciortino}},\
  }\href@noop {} {\bibfield  {journal} {\bibinfo  {journal} {J. Chem. Phys.}\
  }\textbf {\bibinfo {volume} {137}},\ \bibinfo {pages} {084704} (\bibinfo
  {year} {2012})}\BibitemShut {NoStop}%
\bibitem [{\citenamefont {Bernardino}\ and\ \citenamefont {{Telo da
  Gama}}(2012)}]{Bernardino2012}%
  \BibitemOpen
  \bibfield  {author} {\bibinfo {author} {\bibfnamefont {N.~R.}\ \bibnamefont
  {Bernardino}}\ and\ \bibinfo {author} {\bibfnamefont {M.~M.}\ \bibnamefont
  {{Telo da Gama}}},\ }\href@noop {} {\bibfield  {journal} {\bibinfo  {journal}
  {Phys. Rev. Lett.}\ }\textbf {\bibinfo {volume} {109}},\ \bibinfo {pages}
  {116103} (\bibinfo {year} {2012})}\BibitemShut {NoStop}%
\bibitem [{\citenamefont {Dias}\ \emph
  {et~al.}(2013{\natexlab{a}})\citenamefont {Dias}, \citenamefont
  {Ara\'{u}jo},\ and\ \citenamefont {{Telo da Gama}}}]{Dias2013}%
  \BibitemOpen
  \bibfield  {author} {\bibinfo {author} {\bibfnamefont {C.~S.}\ \bibnamefont
  {Dias}}, \bibinfo {author} {\bibfnamefont {N.~A.~M.}\ \bibnamefont
  {Ara\'{u}jo}}, \ and\ \bibinfo {author} {\bibfnamefont {M.~M.}\ \bibnamefont
  {{Telo da Gama}}},\ }\href@noop {} {\bibfield  {journal} {\bibinfo  {journal}
  {Phys. Rev. E}\ }\textbf {\bibinfo {volume} {87}},\ \bibinfo {pages} {032308}
  (\bibinfo {year} {2013}{\natexlab{a}})}\BibitemShut {NoStop}%
\bibitem [{\citenamefont {Dias}\ \emph
  {et~al.}(2013{\natexlab{b}})\citenamefont {Dias}, \citenamefont
  {Ara\'{u}jo},\ and\ \citenamefont {{Telo da Gama}}}]{Dias2013a}%
  \BibitemOpen
  \bibfield  {author} {\bibinfo {author} {\bibfnamefont {C.~S.}\ \bibnamefont
  {Dias}}, \bibinfo {author} {\bibfnamefont {N.~A.~M.}\ \bibnamefont
  {Ara\'{u}jo}}, \ and\ \bibinfo {author} {\bibfnamefont {M.~M.}\ \bibnamefont
  {{Telo da Gama}}},\ }\href@noop {} {\bibfield  {journal} {\bibinfo  {journal}
  {Soft Matter}\ }\textbf {\bibinfo {volume} {9}},\ \bibinfo {pages} {5616}
  (\bibinfo {year} {2013}{\natexlab{b}})}\BibitemShut {NoStop}%
\bibitem [{\citenamefont {Dias}\ \emph
  {et~al.}(2013{\natexlab{c}})\citenamefont {Dias}, \citenamefont
  {Ara\'{u}jo},\ and\ \citenamefont {{Telo da Gama}}}]{Dias2013b}%
  \BibitemOpen
  \bibfield  {author} {\bibinfo {author} {\bibfnamefont {C.~S.}\ \bibnamefont
  {Dias}}, \bibinfo {author} {\bibfnamefont {N.~A.~M.}\ \bibnamefont
  {Ara\'{u}jo}}, \ and\ \bibinfo {author} {\bibfnamefont {M.~M.}\ \bibnamefont
  {{Telo da Gama}}},\ }\href@noop {} {\bibfield  {journal} {\bibinfo  {journal}
  {J. Chem. Phys.}\ }\textbf {\bibinfo {volume} {139}},\ \bibinfo {pages}
  {154903} (\bibinfo {year} {2013}{\natexlab{c}})}\BibitemShut {NoStop}%
\bibitem [{\citenamefont {Shyr}\ \emph {et~al.}(2008)\citenamefont {Shyr},
  \citenamefont {Wernette}, \citenamefont {Wiltzius}, \citenamefont {Lu},\ and\
  \citenamefont {Braun}}]{Shyr2008}%
  \BibitemOpen
  \bibfield  {author} {\bibinfo {author} {\bibfnamefont {M.~H.~S.}\
  \bibnamefont {Shyr}}, \bibinfo {author} {\bibfnamefont {D.~P.}\ \bibnamefont
  {Wernette}}, \bibinfo {author} {\bibfnamefont {P.}~\bibnamefont {Wiltzius}},
  \bibinfo {author} {\bibfnamefont {Y.}~\bibnamefont {Lu}}, \ and\ \bibinfo
  {author} {\bibfnamefont {P.~V.}\ \bibnamefont {Braun}},\ }\href@noop {}
  {\bibfield  {journal} {\bibinfo  {journal} {J. Am. Chem. Soc}\ }\textbf
  {\bibinfo {volume} {130}},\ \bibinfo {pages} {8234} (\bibinfo {year}
  {2008})}\BibitemShut {NoStop}%
\bibitem [{\citenamefont {Pawar}\ and\ \citenamefont
  {Kretzschmar}(2008)}]{Pawar2008}%
  \BibitemOpen
  \bibfield  {author} {\bibinfo {author} {\bibfnamefont {A.~B.}\ \bibnamefont
  {Pawar}}\ and\ \bibinfo {author} {\bibfnamefont {I.}~\bibnamefont
  {Kretzschmar}},\ }\href@noop {} {\bibfield  {journal} {\bibinfo  {journal}
  {Langmuir}\ }\textbf {\bibinfo {volume} {24}},\ \bibinfo {pages} {355}
  (\bibinfo {year} {2008})}\BibitemShut {NoStop}%
\bibitem [{\citenamefont {Tian}\ \emph {et~al.}(2010)\citenamefont {Tian},
  \citenamefont {Cheng}, \citenamefont {Nouvel}, \citenamefont {Geng},\ and\
  \citenamefont {Scherman}}]{Tian2010}%
  \BibitemOpen
  \bibfield  {author} {\bibinfo {author} {\bibfnamefont {F.}~\bibnamefont
  {Tian}}, \bibinfo {author} {\bibfnamefont {N.}~\bibnamefont {Cheng}},
  \bibinfo {author} {\bibfnamefont {N.}~\bibnamefont {Nouvel}}, \bibinfo
  {author} {\bibfnamefont {J.}~\bibnamefont {Geng}}, \ and\ \bibinfo {author}
  {\bibfnamefont {O.~A.}\ \bibnamefont {Scherman}},\ }\href@noop {} {\bibfield
  {journal} {\bibinfo  {journal} {Langmuir}\ }\textbf {\bibinfo {volume}
  {26}},\ \bibinfo {pages} {5323} (\bibinfo {year} {2010})}\BibitemShut
  {NoStop}%
\bibitem [{\citenamefont {Cadilhe}\ \emph {et~al.}(2007)\citenamefont
  {Cadilhe}, \citenamefont {Ara\'{u}jo},\ and\ \citenamefont
  {Privman}}]{Cadilhe2007}%
  \BibitemOpen
  \bibfield  {author} {\bibinfo {author} {\bibfnamefont {A.}~\bibnamefont
  {Cadilhe}}, \bibinfo {author} {\bibfnamefont {N.~A.~M.}\ \bibnamefont
  {Ara\'{u}jo}}, \ and\ \bibinfo {author} {\bibfnamefont {V.}~\bibnamefont
  {Privman}},\ }\href@noop {} {\bibfield  {journal} {\bibinfo  {journal} {J.
  Phys.: Condens. Matter}\ }\textbf {\bibinfo {volume} {19}},\ \bibinfo {pages}
  {065124} (\bibinfo {year} {2007})}\BibitemShut {NoStop}%
\bibitem [{\citenamefont {Ara\'{u}jo}\ \emph {et~al.}(2008)\citenamefont
  {Ara\'{u}jo}, \citenamefont {Cadilhe},\ and\ \citenamefont
  {Privman}}]{Araujo2008}%
  \BibitemOpen
  \bibfield  {author} {\bibinfo {author} {\bibfnamefont {N.~A.~M.}\
  \bibnamefont {Ara\'{u}jo}}, \bibinfo {author} {\bibfnamefont
  {A.}~\bibnamefont {Cadilhe}}, \ and\ \bibinfo {author} {\bibfnamefont
  {V.}~\bibnamefont {Privman}},\ }\href@noop {} {\bibfield  {journal} {\bibinfo
   {journal} {Phys. Rev. E}\ }\textbf {\bibinfo {volume} {77}},\ \bibinfo
  {pages} {031603} (\bibinfo {year} {2008})}\BibitemShut {NoStop}%
\bibitem [{\citenamefont {Einstein}\ and\ \citenamefont
  {Stasevich}(2010)}]{Einstein2010}%
  \BibitemOpen
  \bibfield  {author} {\bibinfo {author} {\bibfnamefont {T.~L.}\ \bibnamefont
  {Einstein}}\ and\ \bibinfo {author} {\bibfnamefont {T.~J.}\ \bibnamefont
  {Stasevich}},\ }\href@noop {} {\bibfield  {journal} {\bibinfo  {journal}
  {Science}\ }\textbf {\bibinfo {volume} {327}},\ \bibinfo {pages} {423}
  (\bibinfo {year} {2010})}\BibitemShut {NoStop}%
\bibitem [{\citenamefont {Doppelbauer}\ \emph {et~al.}(2010)\citenamefont
  {Doppelbauer}, \citenamefont {Bianchi},\ and\ \citenamefont
  {Kahl}}]{Doppelbauer2010}%
  \BibitemOpen
  \bibfield  {author} {\bibinfo {author} {\bibfnamefont {G.}~\bibnamefont
  {Doppelbauer}}, \bibinfo {author} {\bibfnamefont {E.}~\bibnamefont
  {Bianchi}}, \ and\ \bibinfo {author} {\bibfnamefont {G.}~\bibnamefont
  {Kahl}},\ }\href@noop {} {\bibfield  {journal} {\bibinfo  {journal} {J.
  Phys.: Cond. Matter}\ }\textbf {\bibinfo {volume} {22}},\ \bibinfo {pages}
  {104105} (\bibinfo {year} {2010})}\BibitemShut {NoStop}%
\bibitem [{\citenamefont {Marshall}\ and\ \citenamefont
  {Chapman}(2013{\natexlab{a}})}]{Marshall2013}%
  \BibitemOpen
  \bibfield  {author} {\bibinfo {author} {\bibfnamefont {B.~D.}\ \bibnamefont
  {Marshall}}\ and\ \bibinfo {author} {\bibfnamefont {W.~G.}\ \bibnamefont
  {Chapman}},\ }\href@noop {} {\bibfield  {journal} {\bibinfo  {journal} {J.
  Chem. Phys.}\ }\textbf {\bibinfo {volume} {139}},\ \bibinfo {pages} {054902}
  (\bibinfo {year} {2013}{\natexlab{a}})}\BibitemShut {NoStop}%
\bibitem [{\citenamefont {Marshall}\ and\ \citenamefont
  {Chapman}(2013{\natexlab{b}})}]{Marshall2013a}%
  \BibitemOpen
  \bibfield  {author} {\bibinfo {author} {\bibfnamefont {B.~D.}\ \bibnamefont
  {Marshall}}\ and\ \bibinfo {author} {\bibfnamefont {W.~G.}\ \bibnamefont
  {Chapman}},\ }\href@noop {} {\bibfield  {journal} {\bibinfo  {journal} {Phys.
  Rev. E}\ }\textbf {\bibinfo {volume} {87}},\ \bibinfo {pages} {052307}
  (\bibinfo {year} {2013}{\natexlab{b}})}\BibitemShut {NoStop}%
\bibitem [{\citenamefont {Tavares}\ \emph {et~al.}(2014)\citenamefont
  {Tavares}, \citenamefont {Almarza},\ and\ \citenamefont {{Telo da
  Gama}}}]{Tavares2014}%
  \BibitemOpen
  \bibfield  {author} {\bibinfo {author} {\bibfnamefont {J.~M.}\ \bibnamefont
  {Tavares}}, \bibinfo {author} {\bibfnamefont {N.~G.}\ \bibnamefont
  {Almarza}}, \ and\ \bibinfo {author} {\bibfnamefont {M.~M.}\ \bibnamefont
  {{Telo da Gama}}},\ }\href@noop {} {\bibfield  {journal} {\bibinfo  {journal}
  {J. Chem. Phys.}\ }\textbf {\bibinfo {volume} {140}},\ \bibinfo {pages}
  {044905} (\bibinfo {year} {2014})}\BibitemShut {NoStop}%
\bibitem [{\citenamefont {Iwashita}\ and\ \citenamefont
  {Kimura}(2013)}]{Iwashita2013}%
  \BibitemOpen
  \bibfield  {author} {\bibinfo {author} {\bibfnamefont {Y.}~\bibnamefont
  {Iwashita}}\ and\ \bibinfo {author} {\bibfnamefont {Y.}~\bibnamefont
  {Kimura}},\ }\href@noop {} {\bibfield  {journal} {\bibinfo  {journal} {Soft
  Matt.}\ }\textbf {\bibinfo {volume} {9}},\ \bibinfo {pages} {10694} (\bibinfo
  {year} {2013})}\BibitemShut {NoStop}%
\bibitem [{\citenamefont {Iwashita}\ and\ \citenamefont
  {Kimura}(2014)}]{Iwashita2014}%
  \BibitemOpen
  \bibfield  {author} {\bibinfo {author} {\bibfnamefont {Y.}~\bibnamefont
  {Iwashita}}\ and\ \bibinfo {author} {\bibfnamefont {Y.}~\bibnamefont
  {Kimura}},\ }\href@noop {} {\bibfield  {journal} {\bibinfo  {journal} {Soft
  Matt.}\ }\textbf {\bibinfo {volume} {10}},\ \bibinfo {pages} {7170} (\bibinfo
  {year} {2014})}\BibitemShut {NoStop}%
\bibitem [{\citenamefont {Geerts}\ and\ \citenamefont
  {Eiser}(2010)}]{Geerts2010}%
  \BibitemOpen
  \bibfield  {author} {\bibinfo {author} {\bibfnamefont {N.}~\bibnamefont
  {Geerts}}\ and\ \bibinfo {author} {\bibfnamefont {E.}~\bibnamefont {Eiser}},\
  }\href@noop {} {\bibfield  {journal} {\bibinfo  {journal} {Soft Matt.}\
  }\textbf {\bibinfo {volume} {6}},\ \bibinfo {pages} {4647} (\bibinfo {year}
  {2010})}\BibitemShut {NoStop}%
\bibitem [{\citenamefont {Leunissen}\ and\ \citenamefont
  {Frenkel}(2011)}]{Leunissen2011}%
  \BibitemOpen
  \bibfield  {author} {\bibinfo {author} {\bibfnamefont {M.~E.}\ \bibnamefont
  {Leunissen}}\ and\ \bibinfo {author} {\bibfnamefont {D.}~\bibnamefont
  {Frenkel}},\ }\href@noop {} {\bibfield  {journal} {\bibinfo  {journal} {J.
  Chem. Phys.}\ }\textbf {\bibinfo {volume} {134}},\ \bibinfo {pages} {084702}
  (\bibinfo {year} {2011})}\BibitemShut {NoStop}%
\bibitem [{\citenamefont {Kardar}\ \emph {et~al.}(1986)\citenamefont {Kardar},
  \citenamefont {Parisi},\ and\ \citenamefont {Zhang}}]{Kardar1986}%
  \BibitemOpen
  \bibfield  {author} {\bibinfo {author} {\bibfnamefont {M.}~\bibnamefont
  {Kardar}}, \bibinfo {author} {\bibfnamefont {G.}~\bibnamefont {Parisi}}, \
  and\ \bibinfo {author} {\bibfnamefont {Y.-C.}\ \bibnamefont {Zhang}},\
  }\href@noop {} {\bibfield  {journal} {\bibinfo  {journal} {Phys. Rev. Lett.}\
  }\textbf {\bibinfo {volume} {56}},\ \bibinfo {pages} {889} (\bibinfo {year}
  {1986})}\BibitemShut {NoStop}%
\bibitem [{\citenamefont {Yunker}\ \emph
  {et~al.}(2013{\natexlab{a}})\citenamefont {Yunker}, \citenamefont {Lohr},
  \citenamefont {Still}, \citenamefont {Borodin}, \citenamefont {Durian},\ and\
  \citenamefont {Yodh}}]{Yunker2013}%
  \BibitemOpen
  \bibfield  {author} {\bibinfo {author} {\bibfnamefont {P.~J.}\ \bibnamefont
  {Yunker}}, \bibinfo {author} {\bibfnamefont {M.~A.}\ \bibnamefont {Lohr}},
  \bibinfo {author} {\bibfnamefont {T.}~\bibnamefont {Still}}, \bibinfo
  {author} {\bibfnamefont {A.}~\bibnamefont {Borodin}}, \bibinfo {author}
  {\bibfnamefont {D.~J.}\ \bibnamefont {Durian}}, \ and\ \bibinfo {author}
  {\bibfnamefont {A.~G.}\ \bibnamefont {Yodh}},\ }\href@noop {} {\bibfield
  {journal} {\bibinfo  {journal} {Phys. Rev. Lett.}\ }\textbf {\bibinfo
  {volume} {110}},\ \bibinfo {pages} {035501} (\bibinfo {year}
  {2013}{\natexlab{a}})}\BibitemShut {NoStop}%
\bibitem [{\citenamefont {Jensen}\ \emph {et~al.}(1990)\citenamefont {Jensen},
  \citenamefont {Fogedby},\ and\ \citenamefont {Dickman}}]{Jensen1990a}%
  \BibitemOpen
  \bibfield  {author} {\bibinfo {author} {\bibfnamefont {I.}~\bibnamefont
  {Jensen}}, \bibinfo {author} {\bibfnamefont {H.~C.}\ \bibnamefont {Fogedby}},
  \ and\ \bibinfo {author} {\bibfnamefont {R.}~\bibnamefont {Dickman}},\
  }\href@noop {} {\bibfield  {journal} {\bibinfo  {journal} {Phys. Rev. A}\
  }\textbf {\bibinfo {volume} {41}},\ \bibinfo {pages} {3411} (\bibinfo {year}
  {1990})}\BibitemShut {NoStop}%
\bibitem [{\citenamefont {Alencar}\ \emph {et~al.}(1997)\citenamefont
  {Alencar}, \citenamefont {{Andrade Jr.}},\ and\ \citenamefont
  {Lucena}}]{Alencar1997}%
  \BibitemOpen
  \bibfield  {author} {\bibinfo {author} {\bibfnamefont {A.~M.}\ \bibnamefont
  {Alencar}}, \bibinfo {author} {\bibfnamefont {J.~S.}\ \bibnamefont {{Andrade
  Jr.}}}, \ and\ \bibinfo {author} {\bibfnamefont {L.~S.}\ \bibnamefont
  {Lucena}},\ }\href@noop {} {\bibfield  {journal} {\bibinfo  {journal} {Phys.
  Rev. E}\ }\textbf {\bibinfo {volume} {56}},\ \bibinfo {pages} {R2379}
  (\bibinfo {year} {1997})}\BibitemShut {NoStop}%
\bibitem [{\citenamefont {{Aar\~{a}o Reis}}(2002)}]{AaraoReis2002}%
  \BibitemOpen
  \bibfield  {author} {\bibinfo {author} {\bibfnamefont {F.~D.~A.}\
  \bibnamefont {{Aar\~{a}o Reis}}},\ }\href@noop {} {\bibfield  {journal}
  {\bibinfo  {journal} {Phys. Rev. E}\ }\textbf {\bibinfo {volume} {66}},\
  \bibinfo {pages} {027101} (\bibinfo {year} {2002})}\BibitemShut {NoStop}%
\bibitem [{\citenamefont {Henkel}\ \emph {et~al.}(2008)\citenamefont {Henkel},
  \citenamefont {Hinrichsen},\ and\ \citenamefont {L\"{u}beck}}]{Henkel2008b}%
  \BibitemOpen
  \bibfield  {author} {\bibinfo {author} {\bibfnamefont {M.}~\bibnamefont
  {Henkel}}, \bibinfo {author} {\bibfnamefont {H.}~\bibnamefont {Hinrichsen}},
  \ and\ \bibinfo {author} {\bibfnamefont {S.}~\bibnamefont {L\"{u}beck}},\
  }\href@noop {} {\emph {\bibinfo {title} {{Non-equilibrium phase
  transitions}}}}\ (\bibinfo  {publisher} {Springer},\ \bibinfo {address}
  {Bristol},\ \bibinfo {year} {2008})\BibitemShut {NoStop}%
\bibitem [{\citenamefont {Lubeck}(2004)}]{Lubeck2003}%
  \BibitemOpen
  \bibfield  {author} {\bibinfo {author} {\bibfnamefont {S.}~\bibnamefont
  {Lubeck}},\ }\href@noop {} {\bibfield  {journal} {\bibinfo  {journal} {Inter.
  J. Mod. Phys. B}\ }\textbf {\bibinfo {volume} {18}},\ \bibinfo {pages} {3977}
  (\bibinfo {year} {2004})}\BibitemShut {NoStop}%
\bibitem [{\citenamefont {\'{O}dor}(2004)}]{Odor2004}%
  \BibitemOpen
  \bibfield  {author} {\bibinfo {author} {\bibfnamefont {G.}~\bibnamefont
  {\'{O}dor}},\ }\href@noop {} {\bibfield  {journal} {\bibinfo  {journal} {Rev.
  Mod. Phys}\ }\textbf {\bibinfo {volume} {76}},\ \bibinfo {pages} {663}
  (\bibinfo {year} {2004})}\BibitemShut {NoStop}%
\bibitem [{\citenamefont {Loweth}\ \emph {et~al.}(1999)\citenamefont {Loweth},
  \citenamefont {Caldwell}, \citenamefont {Peng}, \citenamefont {Alivisatos},\
  and\ \citenamefont {Schultz}}]{Loweth1999}%
  \BibitemOpen
  \bibfield  {author} {\bibinfo {author} {\bibfnamefont {C.~J.}\ \bibnamefont
  {Loweth}}, \bibinfo {author} {\bibfnamefont {W.~B.}\ \bibnamefont
  {Caldwell}}, \bibinfo {author} {\bibfnamefont {X.}~\bibnamefont {Peng}},
  \bibinfo {author} {\bibfnamefont {A.~P.}\ \bibnamefont {Alivisatos}}, \ and\
  \bibinfo {author} {\bibfnamefont {P.~G.}\ \bibnamefont {Schultz}},\
  }\href@noop {} {\bibfield  {journal} {\bibinfo  {journal} {Angew. Chem. Int.
  Ed.}\ }\textbf {\bibinfo {volume} {38}},\ \bibinfo {pages} {1808} (\bibinfo
  {year} {1999})}\BibitemShut {NoStop}%
\bibitem [{\citenamefont {Largo}\ \emph {et~al.}(2007)\citenamefont {Largo},
  \citenamefont {Tartaglia},\ and\ \citenamefont {Sciortino}}]{Largo2007}%
  \BibitemOpen
  \bibfield  {author} {\bibinfo {author} {\bibfnamefont {J.}~\bibnamefont
  {Largo}}, \bibinfo {author} {\bibfnamefont {P.}~\bibnamefont {Tartaglia}}, \
  and\ \bibinfo {author} {\bibfnamefont {F.}~\bibnamefont {Sciortino}},\
  }\href@noop {} {\bibfield  {journal} {\bibinfo  {journal} {Phys. Rev. E}\
  }\textbf {\bibinfo {volume} {76}},\ \bibinfo {pages} {011402} (\bibinfo
  {year} {2007})}\BibitemShut {NoStop}%
\bibitem [{\citenamefont {Wilber}\ \emph {et~al.}(2007)\citenamefont {Wilber},
  \citenamefont {Doye}, \citenamefont {Louis}, \citenamefont {Noya},
  \citenamefont {Miller},\ and\ \citenamefont {Wong}}]{Wilber2007}%
  \BibitemOpen
  \bibfield  {author} {\bibinfo {author} {\bibfnamefont {A.~W.}\ \bibnamefont
  {Wilber}}, \bibinfo {author} {\bibfnamefont {J.~P.~K.}\ \bibnamefont {Doye}},
  \bibinfo {author} {\bibfnamefont {A.~A.}\ \bibnamefont {Louis}}, \bibinfo
  {author} {\bibfnamefont {E.~G.}\ \bibnamefont {Noya}}, \bibinfo {author}
  {\bibfnamefont {M.~A.}\ \bibnamefont {Miller}}, \ and\ \bibinfo {author}
  {\bibfnamefont {P.}~\bibnamefont {Wong}},\ }\href@noop {} {\bibfield
  {journal} {\bibinfo  {journal} {J. Chem. Phys.}\ }\textbf {\bibinfo {volume}
  {127}},\ \bibinfo {pages} {085106} (\bibinfo {year} {2007})}\BibitemShut
  {NoStop}%
\bibitem [{\citenamefont {L\"{u}beck}(2006)}]{Lubeck2006}%
  \BibitemOpen
  \bibfield  {author} {\bibinfo {author} {\bibfnamefont {S.}~\bibnamefont
  {L\"{u}beck}},\ }\href@noop {} {\bibfield  {journal} {\bibinfo  {journal} {J.
  Stat. Phys.}\ }\textbf {\bibinfo {volume} {123}},\ \bibinfo {pages} {193}
  (\bibinfo {year} {2006})}\BibitemShut {NoStop}%
\bibitem [{\citenamefont {Grassberger}(2006)}]{Grassberger2006}%
  \BibitemOpen
  \bibfield  {author} {\bibinfo {author} {\bibfnamefont {P.}~\bibnamefont
  {Grassberger}},\ }\href@noop {} {\bibfield  {journal} {\bibinfo  {journal}
  {J. Stat. Mech.}\ }\textbf {\bibinfo {volume} {2006}},\ \bibinfo {pages}
  {P01004} (\bibinfo {year} {2006})}\BibitemShut {NoStop}%
\bibitem [{\citenamefont {Yunker}\ \emph
  {et~al.}(2013{\natexlab{b}})\citenamefont {Yunker}, \citenamefont {Durian},\
  and\ \citenamefont {Yodh}}]{Yunker2013b}%
  \BibitemOpen
  \bibfield  {author} {\bibinfo {author} {\bibfnamefont {P.~J.}\ \bibnamefont
  {Yunker}}, \bibinfo {author} {\bibfnamefont {D.~J.}\ \bibnamefont {Durian}},
  \ and\ \bibinfo {author} {\bibfnamefont {A.~G.}\ \bibnamefont {Yodh}},\
  }\href@noop {} {\bibfield  {journal} {\bibinfo  {journal} {Phys. Today}\
  }\textbf {\bibinfo {volume} {66}},\ \bibinfo {pages} {60} (\bibinfo {year}
  {2013}{\natexlab{b}})}\BibitemShut {NoStop}%
\bibitem [{\citenamefont {Yang}\ \emph {et~al.}(2014)\citenamefont {Yang},
  \citenamefont {Li},\ and\ \citenamefont {Sun}}]{Yang2014}%
  \BibitemOpen
  \bibfield  {author} {\bibinfo {author} {\bibfnamefont {X.}~\bibnamefont
  {Yang}}, \bibinfo {author} {\bibfnamefont {C.~Y.}\ \bibnamefont {Li}}, \ and\
  \bibinfo {author} {\bibfnamefont {Y.}~\bibnamefont {Sun}},\ }\href@noop {}
  {\bibfield  {journal} {\bibinfo  {journal} {Soft Matt.}\ }\textbf {\bibinfo
  {volume} {10}},\ \bibinfo {pages} {4458} (\bibinfo {year}
  {2014})}\BibitemShut {NoStop}%
\bibitem [{\citenamefont {Ziff}\ \emph {et~al.}(1986)\citenamefont {Ziff},
  \citenamefont {Gulari},\ and\ \citenamefont {Barshad}}]{Ziff1986}%
  \BibitemOpen
  \bibfield  {author} {\bibinfo {author} {\bibfnamefont {R.~M.}\ \bibnamefont
  {Ziff}}, \bibinfo {author} {\bibfnamefont {E.}~\bibnamefont {Gulari}}, \ and\
  \bibinfo {author} {\bibfnamefont {Y.}~\bibnamefont {Barshad}},\ }\href@noop
  {} {\bibfield  {journal} {\bibinfo  {journal} {Phys. Rev. Lett.}\ }\textbf
  {\bibinfo {volume} {56}},\ \bibinfo {pages} {2553} (\bibinfo {year}
  {1986})}\BibitemShut {NoStop}%
\bibitem [{\citenamefont {Marro}\ and\ \citenamefont
  {Dickman}(1999)}]{Marro1999}%
  \BibitemOpen
  \bibfield  {author} {\bibinfo {author} {\bibfnamefont {J.}~\bibnamefont
  {Marro}}\ and\ \bibinfo {author} {\bibfnamefont {R.}~\bibnamefont
  {Dickman}},\ }\href@noop {} {\emph {\bibinfo {title} {{Nonequilibrium phase
  transitions in lattice models}}}}\ (\bibinfo  {publisher} {Cambridge
  University Press},\ \bibinfo {address} {Cambridge},\ \bibinfo {year}
  {1999})\BibitemShut {NoStop}%
\bibitem [{\citenamefont {Takeuchi}(2014)}]{Takeuchi2014}%
  \BibitemOpen
  \bibfield  {author} {\bibinfo {author} {\bibfnamefont {K.~A.}\ \bibnamefont
  {Takeuchi}},\ }\href@noop {} {\bibfield  {journal} {\bibinfo  {journal} {J.
  Stat. Mech.}\ }\textbf {\bibinfo {volume} {2014}},\ \bibinfo {pages} {P01006}
  (\bibinfo {year} {2014})}\BibitemShut {NoStop}%
\bibitem [{\citenamefont {Takeuchi}\ \emph {et~al.}(2007)\citenamefont
  {Takeuchi}, \citenamefont {Kuroda}, \citenamefont {Chat\'{e}},\ and\
  \citenamefont {Sano}}]{Takeuchi2007}%
  \BibitemOpen
  \bibfield  {author} {\bibinfo {author} {\bibfnamefont {K.~A.}\ \bibnamefont
  {Takeuchi}}, \bibinfo {author} {\bibfnamefont {M.}~\bibnamefont {Kuroda}},
  \bibinfo {author} {\bibfnamefont {H.}~\bibnamefont {Chat\'{e}}}, \ and\
  \bibinfo {author} {\bibfnamefont {M.}~\bibnamefont {Sano}},\ }\href@noop {}
  {\bibfield  {journal} {\bibinfo  {journal} {Phys. Rev. Lett.}\ }\textbf
  {\bibinfo {volume} {99}},\ \bibinfo {pages} {234503} (\bibinfo {year}
  {2007})}\BibitemShut {NoStop}%
\bibitem [{\citenamefont {Shi}\ \emph {et~al.}(2013)\citenamefont {Shi},
  \citenamefont {Avila},\ and\ \citenamefont {Hof}}]{Shi2013}%
  \BibitemOpen
  \bibfield  {author} {\bibinfo {author} {\bibfnamefont {L.}~\bibnamefont
  {Shi}}, \bibinfo {author} {\bibfnamefont {M.}~\bibnamefont {Avila}}, \ and\
  \bibinfo {author} {\bibfnamefont {B.}~\bibnamefont {Hof}},\ }\href@noop {}
  {\bibfield  {journal} {\bibinfo  {journal} {Phys. Rev. Lett.}\ }\textbf
  {\bibinfo {volume} {110}},\ \bibinfo {pages} {204502} (\bibinfo {year}
  {2013})}\BibitemShut {NoStop}%
\bibitem [{\citenamefont {Pine}\ \emph {et~al.}(2005)\citenamefont {Pine},
  \citenamefont {Gollub}, \citenamefont {Brady},\ and\ \citenamefont
  {Leshansky}}]{Pine2005}%
  \BibitemOpen
  \bibfield  {author} {\bibinfo {author} {\bibfnamefont {D.~J.}\ \bibnamefont
  {Pine}}, \bibinfo {author} {\bibfnamefont {J.~P.}\ \bibnamefont {Gollub}},
  \bibinfo {author} {\bibfnamefont {J.~F.}\ \bibnamefont {Brady}}, \ and\
  \bibinfo {author} {\bibfnamefont {A.~M.}\ \bibnamefont {Leshansky}},\
  }\href@noop {} {\bibfield  {journal} {\bibinfo  {journal} {Nature}\ }\textbf
  {\bibinfo {volume} {438}},\ \bibinfo {pages} {997} (\bibinfo {year}
  {2005})}\BibitemShut {NoStop}%
\bibitem [{\citenamefont {Cort\'{e}}\ \emph {et~al.}(2008)\citenamefont
  {Cort\'{e}}, \citenamefont {Chaikin}, \citenamefont {Gollub},\ and\
  \citenamefont {Pine}}]{Corte2008}%
  \BibitemOpen
  \bibfield  {author} {\bibinfo {author} {\bibfnamefont {L.}~\bibnamefont
  {Cort\'{e}}}, \bibinfo {author} {\bibfnamefont {P.~M.}\ \bibnamefont
  {Chaikin}}, \bibinfo {author} {\bibfnamefont {J.~P.}\ \bibnamefont {Gollub}},
  \ and\ \bibinfo {author} {\bibfnamefont {D.~J.}\ \bibnamefont {Pine}},\
  }\href@noop {} {\bibfield  {journal} {\bibinfo  {journal} {Nature Phys.}\
  }\textbf {\bibinfo {volume} {4}},\ \bibinfo {pages} {420} (\bibinfo {year}
  {2008})}\BibitemShut {NoStop}%
\bibitem [{\citenamefont {Hinrichsen}\ \emph {et~al.}(2000)\citenamefont
  {Hinrichsen}, \citenamefont {Livi}, \citenamefont {Mukamel},\ and\
  \citenamefont {Politi}}]{Hinrichsen2000b}%
  \BibitemOpen
  \bibfield  {author} {\bibinfo {author} {\bibfnamefont {H.}~\bibnamefont
  {Hinrichsen}}, \bibinfo {author} {\bibfnamefont {R.}~\bibnamefont {Livi}},
  \bibinfo {author} {\bibfnamefont {D.}~\bibnamefont {Mukamel}}, \ and\
  \bibinfo {author} {\bibfnamefont {A.}~\bibnamefont {Politi}},\ }\href@noop {}
  {\bibfield  {journal} {\bibinfo  {journal} {Phys. Rev. E}\ }\textbf {\bibinfo
  {volume} {61}},\ \bibinfo {pages} {R1032} (\bibinfo {year}
  {2000})}\BibitemShut {NoStop}%
\bibitem [{\citenamefont {Mu\~{n}oz}\ \emph {et~al.}(2005)\citenamefont
  {Mu\~{n}oz}, \citenamefont {de~los Santos},\ and\ \citenamefont {{Telo da
  Gama}}}]{Muoz2005}%
  \BibitemOpen
  \bibfield  {author} {\bibinfo {author} {\bibfnamefont {M.~A.}\ \bibnamefont
  {Mu\~{n}oz}}, \bibinfo {author} {\bibfnamefont {F.}~\bibnamefont {de~los
  Santos}}, \ and\ \bibinfo {author} {\bibfnamefont {M.~M.}\ \bibnamefont
  {{Telo da Gama}}},\ }\href@noop {} {\bibfield  {journal} {\bibinfo  {journal}
  {Eur. Phys. J. B}\ }\textbf {\bibinfo {volume} {43}},\ \bibinfo {pages} {73}
  (\bibinfo {year} {2005})}\BibitemShut {NoStop}%
\bibitem [{\citenamefont {Ginelli}\ \emph {et~al.}(2003)\citenamefont
  {Ginelli}, \citenamefont {Ahlers}, \citenamefont {Livi}, \citenamefont
  {Mukamel}, \citenamefont {Pikovsky}, \citenamefont {Politi},\ and\
  \citenamefont {Torcini}}]{Ginelli2003}%
  \BibitemOpen
  \bibfield  {author} {\bibinfo {author} {\bibfnamefont {F.}~\bibnamefont
  {Ginelli}}, \bibinfo {author} {\bibfnamefont {V.}~\bibnamefont {Ahlers}},
  \bibinfo {author} {\bibfnamefont {R.}~\bibnamefont {Livi}}, \bibinfo {author}
  {\bibfnamefont {D.}~\bibnamefont {Mukamel}}, \bibinfo {author} {\bibfnamefont
  {A.}~\bibnamefont {Pikovsky}}, \bibinfo {author} {\bibfnamefont
  {A.}~\bibnamefont {Politi}}, \ and\ \bibinfo {author} {\bibfnamefont
  {A.}~\bibnamefont {Torcini}},\ }\href@noop {} {\bibfield  {journal} {\bibinfo
   {journal} {Phys. Rev. E}\ }\textbf {\bibinfo {volume} {68}},\ \bibinfo
  {pages} {065102(R)} (\bibinfo {year} {2003})}\BibitemShut {NoStop}%
\bibitem [{\citenamefont {de~los Santos}\ \emph {et~al.}(2002)\citenamefont
  {de~los Santos}, \citenamefont {{Telo da Gama}},\ and\ \citenamefont
  {Mu\~{n}oz}}]{DelosSantos2002}%
  \BibitemOpen
  \bibfield  {author} {\bibinfo {author} {\bibfnamefont {F.}~\bibnamefont
  {de~los Santos}}, \bibinfo {author} {\bibfnamefont {M.~M.}\ \bibnamefont
  {{Telo da Gama}}}, \ and\ \bibinfo {author} {\bibfnamefont {M.~A.}\
  \bibnamefont {Mu\~{n}oz}},\ }\href@noop {} {\bibfield  {journal} {\bibinfo
  {journal} {Europhys. Lett.}\ }\textbf {\bibinfo {volume} {57}},\ \bibinfo
  {pages} {803} (\bibinfo {year} {2002})}\BibitemShut {NoStop}%
\bibitem [{\citenamefont {de~los Santos}\ \emph {et~al.}(2003)\citenamefont
  {de~los Santos}, \citenamefont {{Telo da Gama}},\ and\ \citenamefont
  {Mu\~{n}oz}}]{DelosSantos2003}%
  \BibitemOpen
  \bibfield  {author} {\bibinfo {author} {\bibfnamefont {F.}~\bibnamefont
  {de~los Santos}}, \bibinfo {author} {\bibfnamefont {M.~M.}\ \bibnamefont
  {{Telo da Gama}}}, \ and\ \bibinfo {author} {\bibfnamefont {M.~A.}\
  \bibnamefont {Mu\~{n}oz}},\ }\href@noop {} {\bibfield  {journal} {\bibinfo
  {journal} {Phys. Rev. E}\ }\textbf {\bibinfo {volume} {67}},\ \bibinfo
  {pages} {021607} (\bibinfo {year} {2003})}\BibitemShut {NoStop}%
\end{thebibliography}%

\end{document}